\documentclass[journal]{IEEEtran}

\newcommand{\norm}[1]{\left\lVert#1\right\rVert}

\usepackage{enumerate}

\usepackage{amsfonts,amssymb,amsmath}
\DeclareMathOperator{\Tr}{Tr}

\usepackage{bm,bbm}
\usepackage{subfig}
\usepackage{array}
\usepackage{breqn}

\usepackage{booktabs}
\usepackage{siunitx}

\usepackage{adjustbox}
\usepackage{multirow}
\usepackage{color}
\usepackage[noabbrev]{cleveref}
\usepackage{cite}

\usepackage{times}
\usepackage{multicol}
\usepackage{graphicx}

\usepackage{siunitx}

\begin{document}
%
\title{A PolSAR Scattering Power Factorization Framework and Novel Roll-Invariant Parameters Based Unsupervised Classification Scheme Using a Geodesic Distance}

\author{Debanshu~Ratha,~\IEEEmembership{Student Member,~IEEE,}~Eric~Pottier,~\IEEEmembership{Fellow,~IEEE}\\ 
Avik~Bhattacharya,~\IEEEmembership{Senior Member,~IEEE,}~and~Alejandro~C.~Frery,~\IEEEmembership{Senior Member,~IEEE}
\thanks{D.~Ratha and A.~Bhattacharya are with Centre of Studies in Resources Engineering, Indian Institute of Technology Bombay, India - 400076. (e-mail: debanshu.ratha@gmail.com, avikb@csre.iitb.ac.in)}
\thanks{E.~Pottier is with the Institut d'\'Electronique et de T\'el\'ecommunications de Rennes (I.E.T.R)
	UMR CNRS 6164, Universit\'e Rennes 1, France. (e-mail: eric.pottier@univ-rennes1.fr)}
\thanks{Alejandro~C.~Frery is with the Laborat\'orio de Computa\c c\~ao Cient\'ifica e An\'alise Num\'erica, Universidade Federal de Alagoas, Macei\'o, Brazil. (e-mail: acfrery@laccan.ufal.br)}
}

\markboth{IEEE Trans.\ Geosc.\ Rem.\ Sens.,~Vol.~XX, No.~XX, Month~XXXX}%
{Shell \MakeLowercase{\textit{Ratha et al.}}: PolSAR Scattering Power Factorization with a Geodesic Distance}

\IEEEtitleabstractindextext{%
\begin{abstract}
We propose a generic Scattering Power Factorization Framework (SPFF) for Polarimetric Synthetic Aperture Radar (PolSAR) data to directly obtain $N$ scattering power components along with a residue power component for each pixel. 
Each scattering power component is factorized into similarity (or dissimilarity) using elementary targets and a generalized random volume model. 
The similarity measure is derived using a geodesic distance between pairs of $4\times4$ real Kennaugh matrices. 
In standard model-based decomposition schemes, the $3\times3$ Hermitian positive semi-definite covariance (or coherency) matrix is expressed as a weighted linear combination of scattering targets following a fixed hierarchical process. 
In contrast, under the proposed framework, a convex splitting of unity is performed to obtain the weights while preserving the dominance of the scattering components. 
The product of the total power ($\text{Span}$) with these weights provides the non-negative scattering power components. 
Furthermore, the framework along the geodesic distance is effectively used to obtain specific roll-invariant parameters which are then utilized to design an unsupervised classification scheme. 
The SPFF, the roll invariant parameters, and the classification results are assessed using C-band RADARSAT-2 and L-band ALOS-2 images of San Francisco. 
\end{abstract}

\begin{IEEEkeywords}
PolSAR, 
Scattering Power, 
Factorization, 
Framework, 
Geodesic Distance, 
Roll-Invariant parameters, 
Unsupervised Classification, 
Radar Polarimetry 
\end{IEEEkeywords}}

\maketitle

\IEEEdisplaynontitleabstractindextext
\IEEEpeerreviewmaketitle

\IEEEPARstart{T}{arget} decomposition (TD) theorems are an essential avenue of research in the study of Polarimetric Synthetic Aperture Radar (PolSAR) imagery. In this context, the study of light scattering by small anisotropic particles by Chandrasekhar~\cite{chandrasekhar2013radiative} was the first instance of a TD. 
Later, Huynen~\cite{Huynen_1970_PhD_Thesis} rigorously formulated this notion and laid the foundations of the modern day TDs.  

According to Huynen, the objective of TD is to identify the average scattering mechanism within the pixel in the form of a rank-1 covariance/coherency matrix. 
The interpretation of this information is achieved by obtaining a set of unique parameters often roll-invariant in nature for a description of the target under study. 
On the one hand, this approach is utilized in the eigenvalue/eigenvector based decomposition schemes. 
On the other hand, model-based decompositions interpret the observation as a weighted linear combination of specific scattering mechanisms. 
In the later, the scattering components can be rank-1 or distributed (rank $\geq 1$, e.g. volume scattering models) targets~\cite{lee2009polarimetric}. 

The extraction of a desirable rank-1 (pure or coherent) target is often synonymous with the most dominant scattering mechanism component from the observation~\cite{Cloude85}. 
However, retaining all the components (in decreasing order of dominance) is more useful for a complete and more realistic characterization of the target. 
This task is leveraged for better interpretation of the observation by model-based decompositions~\cite{Freeman98, Yamaguchi2005}, and for uniqueness, by the eigenvalue-eigenvector based decompositions~\cite{Cloude1992, Touzi2007}. 

Recently, Xu et al.~\cite{Xu2017} brought together rank-1 PolSAR decomposition, model-based decomposition, and image clustering under the single umbrella of image factorization problems. 
Motivated by the concept of factorization, Ratha et al.~\cite{Ratha2018Fact} proposed an alternative approach to characterize the observation through a vector of bounded distances from elementary scatterers. 

In this context, we utilized the geodesic distance ($GD$) measured over the unit sphere centered at the origin in the space of real $4 \times 4$ matrices. 
This unit sphere contains all the normalized Kennaugh ($\mathbf{K}$) matrices which are equivalent to the second-order information conveyed by the coherency (or covariance) matrices corresponding to an observation. 
On the unit sphere, the proximity of the observation to the respective elementary scatterer determines the order of the dominant scattering mechanism.

Often the validity of TDs is assessed based on 
(a)~the preservation of dominance order of scattering mechanism components, 
(b)~the reduction in the number of pixels with negative power, and 
(c)~the conservation of total power ($\text{Span}$). 
The usability is evaluated based on better scattering mechanism discrimination, classification, and accurate biophysical parameter extraction~\cite{Chen2014}.         

%
In this perspective, we propose a novel generalized and flexible scattering power factorization (SPFF) framework using a similarity measure derived from the geometrically motivated geodesic distance. 
At first, a convex splitting of unity is performed to obtain the weights preserving the order of dominance of scattering components. 
The individual weights stem from the similarity and dissimilarity (distance) of the observed pixel with the input scattering models. 
Finally, the multiplication of the total power ($\text{Span}$) with these weights provides the non-negative scattering power components. 
In the process, we also obtain some novel roll-invariant parameters which we use for designing a new unsupervised classification scheme for PolSAR images.

Actually, SPFF should not be categorized as a TD, because we split the $\text{Span}$ instead of the matrix representation of the full observation. 
However, at the same time, the criteria of validity and the assessment of SPFF coincides with that of TDs and hence, the comparison. 

In the manuscript, 
Section~\ref{Sec:GD} discusses the formulation of geodesic distance (on the unit sphere), and its particular advantages when working with PolSAR data.
Section~\ref{Sec:DataSets} provides the details about the data sets used for the results. Section~\ref{Sec:RIP} is devoted to the novel roll-invariant parameters which are a function of the geodesic distance of the observation from roll-invariant targets. These parameters are compared against standard roll-invariant parameters existing in PolSAR literature. Section~\ref{Sec:Zone} discusses the identification of scattering zones using the proposed roll-invariant parameters. In Section~\ref{Sec:Classification} we propose an alternative unsupervised classification scheme for PolSAR images. Section~\ref{Sec:SPFF} describes the general flowchart showing the steps involved in the SPFF. In Section~\ref{Sec:SPFF} the SPFF is utilized for multi-looked PolSAR images using a generalized volume scattering model~\cite{Antropov2011}. 
The results are discussed in Section~\ref{Sec:Results}. Section~\ref{Sec:Conclusion} concludes this manuscript. 
    
%
%

\section{Geodesic Distance on the Unit Sphere}\label{Sec:GD}

The geometry of the unit sphere plays an important role in polarimetry theory. 
For example, the Poincar\'e sphere is used to visualize the state of polarization~\cite{lee2009polarimetric} in the Stokes formalism.
Similarly, the unit sphere in the space of $4\times4$ real matrices is ideal for studying scattering behavior of targets
in the context of Kennaugh matrices.

In this case, the geodesic distance~\cite{nakahara2003geometry} on the unit sphere is a natural way to measure the dissimilarity between the targets. 
In particular, the geodesic distance ($GD$) on the unit sphere of $4 \times 4$ real matrices was found to be useful for several applications such as change detection~\cite{Ratha_CD_2017}, unsupervised land-cover classification~\cite{Ratha_SM_2017}, vegetation monitoring~\cite{Ratha_GRVI_2019} and extraction of urban footprint~\cite{Ratha_RBUI_2019} using PolSAR images. 
The $GD$ between two Kennaugh matrices $\mathbf{K}_1$ and $\mathbf{K}_2$ is defined as 
\begin{equation}
GD(\mathbf{K}_1,\mathbf{K}_2) =  \frac{2}{\pi} \cos^{-1}\frac{\Tr(\mathbf{K}_1^T\mathbf{K}_2)}{\sqrt{\Tr(\mathbf{K}_1^T\mathbf{K}_1)}\sqrt{\Tr(\mathbf{K}_2^T\mathbf{K}_2)}} .
\label{eq:GD_Ken}
\end{equation}
Under this definition, $GD$ is the distance between the projections of $\mathbf{K}_1$ and $\mathbf{K}_2$ on the unit sphere centered at the origin in the space of $4 \times 4$ real matrices.   

In the following section, we explore the properties of $GD$ and provide its physical significance.

\subsection{Properties of GD and PolSAR significance}\label{Ps}
Here we discuss a few more properties of $GD$ relevant to PolSAR image analysis and interpretation.

\begin{enumerate}[(P1)]
    
    \item By definition, $GD$ is bounded in $[0,1]$.  
    
    \emph{Significance:} Hence, $GD$ is desirable over unbounded metrics, e.g., the Euclidean distance. Moreover, it can be directly used as an element of a feature vector for algorithms designed for PolSAR applications.
       
    \item $GD$ is scale invariant i.e., $GD(\lambda_1\mathbf{K}_1,\lambda_2\mathbf{K}_2) = GD(\mathbf{K}_1,\mathbf{K}_2)$ where $\lambda_1$ and $\lambda_2$ are positive real numbers.
    
    \emph{Significance:} This property makes $GD$ a metric that captures changes in scattering mechanism only, as it is invariant under the scaling by the $\text{Span}$; identical targets with unequal power return can still be identified as one. 
    
    
    \item $GD$ is invariant under orthogonal transformation of the basis i.e., for $\mathbf{P}\in \mathbb{O}(4)$ (the set of $4\times4$ orthogonal matrices such that $\mathbf{P}^{T}\mathbf{P}=\mathbf{P}\mathbf{P}^T = \mathbf{I}$),
    \begin{equation*}
    GD(\mathbf{P}^{T}\mathbf{K}_1\mathbf{P},\mathbf{P}^{T}\mathbf{K}_2\mathbf{P}) = GD(\mathbf{K}_1,\mathbf{K}_2).
    \end{equation*}
    
    \emph{Significance:} In polarimetry there exists special matrices $O_4(2\phi)$, $O_4(2\tau)$ and $O_4(2\alpha)$. 
    These are used in combination i.e., $O_4(2\phi,2\tau,2\alpha) = O_4(2\phi)O(2\tau)O_4(2\alpha)$ for changing the polarimetric basis from one orthonormal system to another in the Kennaugh matrix representation~\cite{lee2009polarimetric}. 
    These matrices are orthogonal, hence, (P3) makes $GD$ invariant under the orthogonal transformation of wave polarization basis.
\end{enumerate}

\subsection{GD for other PolSAR data representations}

The current form of $GD$ seems to have a restrictive definition, which is suitable only for the Kennaugh matrices. 
In this section, we work out its equivalent forms for the covariance, coherency and the scattering matrices.

Let us recall the conversions from $\mathbf{S}$ and $\mathbf{T}$ to $\mathbf{K}$. Given a scattering matrix $\mathbf{S}$, the $4 \times 4$ real Kennaugh matrix $\mathbf{K}$ is defined as~\cite{lee2009polarimetric}:
\begin{equation}\label{coKen}
\mathbf{K} = \frac{1}{2}\mathbf{A}^*(\mathbf{S} \otimes \mathbf{S}^*) \mathbf{A}^{*T}, \quad \mathbf{A} = \left[\begin{array}{cccc}
1 & 0 & 0 & 1\\
1 & 0 & 0 & -1\\
0 & 1 & 1 & 0\\
0 & j & -j & 0
\end{array}\right],
\end{equation}
where $\otimes$ is the Kronecker product, superscripts $*$ and $T$ denote conjugate and transpose respectively, and  $j = \sqrt{-1}$. Alternatively, the Kennaugh matrix for the incoherent case can be obtained from the coherency matrix $\mathbf{T}$ as follows~\cite{cloude2010polarisation}:
\begin{equation}\label{incoKen}
\mathbf{K} =\resizebox{0.8\hsize}{!}{$
    \left[\begin{array}{cccc}
    \frac{T_{11}+T_{22}+ T_{33}}{2} & \Re(T_{12}) & \Re(T_{13}) & \Im(T_{23})\\
    \Re(T_{12}) & \frac{T_{11}+T_{22}-T_{33}}{2} & \Re(T_{23}) & \Im(T_{13})\\
    \Re(T_{13}) & \Re(T_{23}) & \frac{T_{11}-T_{22}+T_{33}}{2} & - \Im(T_{12})\\
    \Im(T_{23}) & \Im(T_{13}) & - \Im(T_{12}) &\frac{-T_{11}+T_{22}+T_{33}}{2}
    \end{array}\right],
    $}
\end{equation}
where $\Re(\cdot)$ and $\Im(\cdot)$ denote real and imaginary parts of a complex number.

Let $\mathbf{K}$ and $\bar{\mathbf{K}}$ be two Kennaugh matrices and $\mathbf{T}$ and $\bar{\mathbf{T}}$ be their corresponding coherency matrices. 

Note that the expression $\Tr({\mathbf{K}^T\bar{\mathbf{K}}})$ is the scalar dot product of real matrices $\mathbf{K}$ and $\bar{\mathbf{K}}$. Thus, the sum of product of diagonal and non-diagonal elements can be separated from its expression i.e.,

\begin{eqnarray}
\Tr({\mathbf{K}^T\bar{\mathbf{K}}}) & = & \sum_{i=1}^{4}\sum_{j=1}^{4}K_{ij}\bar{K}_{ij}\\
& = & \sum_{i=1}^{4} K_{ii}\bar{K}_{ii} + \sum_{\substack{i = 1\\i\neq j}}^{4}\sum_{j=1}^{4}K_{ij}\bar{K}_{ij}
\end{eqnarray}

By substituting the relationship between entries of Kennaugh and corresponding coherency matrix~\eqref{incoKen} into the diagonal and off-diagonal parts in the preceding equation we obtain upon simplification,
\begin{eqnarray}
\sum_{i=1}^{4} K_{ii}\bar{K}_{ii} & = & \sum_{i=1}^{4}T_{ii}\bar{T}_{ii}\\
\sum_{\substack{i = 1\\i\neq j}}^{4}\sum_{j=1}^{4}K_{ij}\bar{K}_{ij} & = &  \sum_{\substack{i=1\\i<j}}^{4}\sum_{j=1}^{4} 2(\Re(T_{ij})\Re(\bar{T}_{ij})\dots\nonumber\\
&  &\dots+\Im(T_{ij})\Im(\bar{T}_{ij}))\\
& = & \sum_{\substack{i=1\\i\neq j}}^{4}\sum_{j=1}^{4}\Re(T_{ij})\Re(\bar{T}_{ij})\dots\nonumber\\
&  &\dots+\Im(T_{ij})\Im(\bar{T}_{ij}).
\end{eqnarray}

The last expression is obtained due to the hermitian nature of the coherency matrices i.e., $T_{ji} = T_{ij}^*$ and $\bar{T}_{ji} = \bar{T}_{ij}^*$. Note that this property forces the diagonal elements of coherency matrices $T_{ii}$ and $\bar{T}_{ii}$ to be real numbers.

Now we expand the term $\Tr(\mathbf{T}^H\bar{\mathbf{T}})$ which the scalar dot product for complex matrices and obtain the following expression,
\begin{eqnarray}
\Tr(\mathbf{T}^H\bar{\mathbf{T}}) & = & \sum_{i=1}^{4}T_{ii}\bar{T}_{ii} + \sum_{\substack{i=1 \\i\neq j}}^{4}\sum_{j=1}^{4}\Re(T_{ij})\Re(\bar{T}_{ij})\dots\nonumber\\
&  &\dots+\Im(T_{ij})\Im(\bar{T}_{ij})\nonumber\\
& = & \Tr({\mathbf{K}^T\bar{\mathbf{K}}}).
\label{dotT2K}
\end{eqnarray}

It is interesting to note that the scalar dot product for complex coherency matrices maps to the real numbers due to its equality with scalar dot product for real symmetric Kennaugh matrices.

Using the identity obtained in~\eqref{dotT2K} we further obtain,  
    \begin{equation}\label{KT}
    \frac{\Tr(\mathbf{K}_1^T\mathbf{K}_2)}{\sqrt{\Tr(\mathbf{K}_1^T\mathbf{K}_1)}\sqrt{\Tr(\mathbf{K}_2^T\mathbf{K}_2)}} =  \frac{\Tr(\mathbf{T}_1^H\mathbf{T}_2)}{\sqrt{\Tr(\mathbf{T}_1^H\mathbf{T}_1)}\sqrt{\Tr(\mathbf{T}_2^H\mathbf{T}_2)}},
    \end{equation}
    where the superscript $H$ stands for the conjugate transpose of the matrix. 
    
    The relationship between a covariance matrix $\mathbf{C}$ and the corresponding coherency matrix $\mathbf{T}$ is expressed via a special unitary matrix $\mathbf{U}_{3(L\mapsto P)} \in \mathbb{SU}(3) = \{\mathbf{U} \in \mathbb{C}^3 \times \mathbb{C}^3:\mathbf{U}^H \mathbf{U} = \mathbf{I} = \mathbf{U} \mathbf{U}^H \mbox{ and } \det(\mathbf{U})=1\}$ where $\mathbf{I}$ is the $3 \times 3$ identity matrix and $\det(\cdot)$ denotes the determinant of the matrix. Then the transformation from $\mathbf{C}$ to $\mathbf{T}$ is obtained as follows:
\begin{equation}\label{eq:lp}
\mathbf{T} = \mathbf{U}_{3(L\mapsto P)} \mathbf{C} \mathbf{U}_{3(L\mapsto P)}^H ,
\end{equation}
where
\begin{equation}
\mathbf{U}_{3(L\mapsto P)} = \frac{1}{\sqrt{2}}
\left[\begin{array}{ccc}
1 & 0 & 1\\
1 & 0 & -1\\
0 & \sqrt{2} & 0
\end{array}\right],
\end{equation}
and, by definition, satisfying the property:
\begin{equation}
\mathbf{U}_{3(L\mapsto P)} \mathbf{U}_{3(L\mapsto P)}^H = \mathbf{U}_{3(L\mapsto P)}^H \mathbf{U}_{3(L\mapsto P)} = \mathbf{I}. 
\end{equation}
Substituting~\eqref{eq:lp} in~\eqref{KT} we obtain,
\begin{equation}\label{eq:TC}
\frac{\Tr(\mathbf{T}_1^H\mathbf{T}_2)}{\sqrt{\Tr(\mathbf{T}_1^H\mathbf{T}_1)}\sqrt{\Tr(\mathbf{T}_2^H\mathbf{T}_2)}} = \frac{\Tr(\mathbf{C}_1^H\mathbf{C}_2)}{\sqrt{\Tr(\mathbf{C}_1^H\mathbf{C}_1)}\sqrt{\Tr(\mathbf{C}_2^H\mathbf{C}_2)}}.
\end{equation}
Thus, we can obtain $GD$ from $\mathbf{C}$ or $\mathbf{T}$~\eqref{eq:TC} interchangeably.

Under coherent conditions, where $\mathbf{K}_1$ and $\mathbf{K}_2$ are derived from corresponding scattering matrices $\mathbf{S}_1$ and $\mathbf{S}_2$ using~\eqref{coKen}, the numerator term in~\eqref{KT} can be simplified as follows,
\begin{eqnarray*}\label{dotK2S}
	& & \Tr(\mathbf{K}_1^T\mathbf{K}_2)\\
	& = & \Tr(\mathbf{K}_1^H\mathbf{K}_2)\\
	& = & \Tr((\frac{1}{2}\mathbf{A}^*(\mathbf{S}_1 \otimes \mathbf{S}_1^*)\mathbf{A}^H)^H\frac{1}{2}\mathbf{A}^*(\mathbf{S}_2 \otimes \mathbf{S}_2^*) \mathbf{A}^H)\\
	&=& \frac{1}{4}\Tr(\mathbf{A}(\mathbf{S}_1 \otimes \mathbf{S}_1^*)^H\mathbf{A}^T\mathbf{A}^*(\mathbf{S}_2 \otimes \mathbf{S}_2^*) \mathbf{A}^H)\\
	&=& \frac{1}{4}\Tr(\mathbf{A}(\mathbf{S}_1 \otimes \mathbf{S}_1^*)^H(\mathbf{S}_2 \otimes \mathbf{S}_2^*) \mathbf{A}^H)\\
	&=& \frac{1}{4}\Tr((\mathbf{S}_1 \otimes \mathbf{S}_1^*)^H(\mathbf{S}_2 \otimes \mathbf{S}_2^*) \mathbf{A}^H\mathbf{A})\\
	&=& \frac{1}{4}\Tr((\mathbf{S}_1 \otimes \mathbf{S}_1^*)^H(\mathbf{S}_2 \otimes \mathbf{S}_2^*)).
\end{eqnarray*}
To arrive at this final form, we have utilized in order: the real nature of Kennaugh matrices, the definition of $H = *T = T*$, the property of $\Tr$ being invariant under cyclic permutation of its arguments and lastly, the identity $\mathbf{A}^H\mathbf{A}=\mathbf{I}$.  
Thus, the equivalent expression using $\mathbf{S}_1$ and $\mathbf{S}_2$ corresponding to the identities~\eqref{KT} and~\eqref{eq:TC} is given as,
\begin{equation}\label{KS}
\resizebox{0.91\hsize}{!}{%
    $\frac{\Tr((\mathbf{S}_1\otimes\mathbf{S}_1^*)^H(\mathbf{S}_2\otimes\mathbf{S}_2^*))}{\sqrt{\Tr((\mathbf{S}_1\otimes\mathbf{S}_1^*)^H(\mathbf{S}_1\otimes\mathbf{S}_1^*))}\sqrt{\Tr((\mathbf{S}_2\otimes\mathbf{S}_2^*)^H(\mathbf{S}_2\otimes\mathbf{S}_2^*))}}$}.
\end{equation}
Thus, when this is substituted in~\eqref{eq:GD_Ken}, provides a way to obtain the $GD$ in terms of the scattering matrices.

In the case of a coherent target, the denominator can be further simplified. 
It can be derived from the definition of the covariance matrix $\mathbf{C}$ without the ensemble averaging. 
For coherent monostatic full polarimetric SAR measurements, the $\text{Span}$ is defined as: 
\begin{eqnarray}
\text{Span} & = & |S_{HH}|^2 + 2|S_{HV}|^2 + |S_{VV}|^2\\
& = & \sqrt{\Tr(\mathbf{C}^H\mathbf{C})} \\
& = & \sqrt{\Tr(\mathbf{T}^H\mathbf{T})} \\
& = & \sqrt{\Tr(\mathbf{K}^T\mathbf{K})}  .
\end{eqnarray}
Expanding the last equality we obtain,
\begin{eqnarray}
(\text{Span})^2 & = & \Tr(\mathbf{K}^T\mathbf{K})\\
\Rightarrow 4K_{11}^2 & = & \sum_{i=1}^{4}\sum_{j=1}^{4} K_{ij}^2 , \label{FWE}
\end{eqnarray}
which is precisely the Fry-Kattawar equation~\cite{Fry81} initially derived for the Stokes matrix in optical polarimetry. 
Many of the equalities that are discussed in~\cite{ Fry81} were first explicitly given by~\cite{abhy69}, then revised and extended in~\cite{Fry81}. 
At the same time, this equality is a necessary but not sufficient condition to warrant that $\mathbf{K}$ is derived from a $\mathbf{S}$ matrix~\cite{cloude2010polarisation}. 
Thus, we have obtained the formulation for $GD$ for all the data representations in PolSAR.  

Although we have a distance in the form of $GD$, it would be better to construct a measure of similarity from it. 
This formulation can be achieved by complementing it with the unit, i.e.,
\begin{equation}
f_{\text{ref}} = 1 - GD(\mathbf{K},\mathbf{K}_{\text{ref}}),
\label{ssim}
\end{equation} 
where $\mathbf{K}$ is an observed Kennaugh matrix and $\mathbf{K}_{\text{ref}}$ is the reference elementary scatterers. 
In this sense, $f_{\text{ref}}$ is a similarity and the corresponding $GD$ is a dissimilarity. 
In PolSAR literature, Yang et al.~\cite{Yang2001Similarity}, Touzi and Charboneau~\cite{Touzi2002}, and Chen et al.~\cite{Chen2013Similarity} discuss similarity-based approaches for describing scattering phenomenon from PolSAR images.    

Thus, the $GD$ is advantageous in terms of its physical significance with parallel definitions across all data representations in PolSAR. Its simple form makes it ideal for computational implementation in several PolSAR applications~\cite{Ratha_CD_2017,Ratha_SM_2017,Ratha_GRVI_2019,Ratha_RBUI_2019}. 

\section{Data Sets}\label{Sec:DataSets}

We have utilized two PolSAR images of the San Francisco (SF) Area. 
The first one is a C-Band RADARSAT-2 (RS-2) acquired on 9th April 2008. 
The near to far range incidence angle is specified as \SIrange{28.02}{29.82}{\degree}. 
The original image is multi-looked by a factor of $2$ in range and $4$ in the azimuth resulting in a \SI{20}{m} ground resolution. 

\begin{figure}[htb]
	\centering
	\subfloat[Pauli RGB 1 ]{\includegraphics[width=0.39\columnwidth]{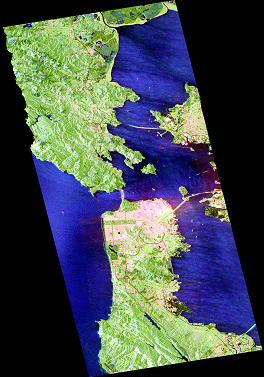}\label{subfig:PauliRGB_RS2_SF}}\quad
	\subfloat[Pauli RGB 2]{\includegraphics[width=0.39\columnwidth]{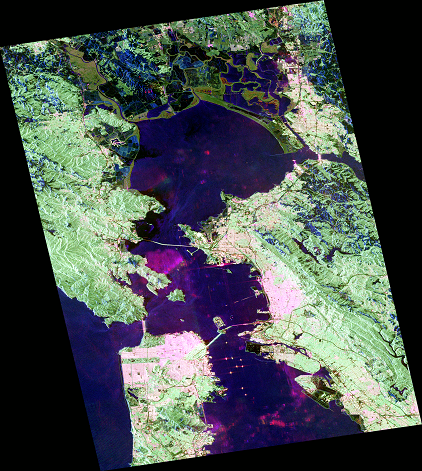}\label{subfig:PauliRGB_ALOS2_SF}}        
	\caption{Pauli RGB images of RS-2 C-band (on left) and ALOS-2 L-band (on right) acquisition over San Francisco.}
	\label{fig:Dataset_SF}
\end{figure}

The other image is a L-Band ALOS-2 acquisition on 29th January 2019. 
The off-nadir angle is specified as \SI{30.8}{\degree}. 
The original image is multi-looked by a factor of $3$ in range and $5$ in the azimuth resulting in a \SI{15.7}{m} ground resolution. 
Fig.~\ref{fig:Dataset_SF} shows the two Pauli RGBs for these data sets.

\section{New Roll Invariant Parameters}\label{Sec:RIP}
In the phenomenon of a roll, the antenna coordinate system is rotated by an angle $\theta$ about the radar line of sight (LoS)~\cite{lee2009polarimetric}. In such a case, the observed Kennaugh matrix $\mathbf{K}$ transforms as follows,
\begin{equation}\label{eq:roll_K}
\mathbf{K}(\theta) = \mathbf{R}(\theta)\mathbf{K}\mathbf{R}(\theta)^T
\end{equation}
where the (orthogonal) rotation matrix $\mathbf{R}(\theta)$ is given by
\begin{equation}
\mathbf{R}(\theta) =\left[\begin{array}{cccc}
1 & 0  & 0 & 0\\
0 & \cos{2\theta} & -\sin{2\theta} & 0\\
0 & \sin{2\theta} &  \cos{2\theta} & 0\\\
0 & 0 & 0 & 1	  
\end{array}\right].
\end{equation}
Let $\mathbf{K}_0$ be the Kennaugh matrix for a  roll-invariant target. A roll-invariant target has the property of preserving its scattering signature despite a roll i.e.,
\begin{equation}\label{eq:rinvK}
\mathbf{R}(\theta)\mathbf{K}_0\mathbf{R}(\theta)^T = \mathbf{K}_0
\end{equation}
for any value of the $\theta$ angle.
Thus, the geodesic distance between $\mathbf{K}(\theta)$ and the roll invariant target $\mathbf{K}_0$ can be further simplified in the following way,
\begin{equation}
GD(\mathbf{K}(\theta),\mathbf{K}_0) = GD(\mathbf{K}(\theta),\mathbf{K}_0(\theta)) = GD(\mathbf{K},\mathbf{K}_0).
\end{equation}
The first step was obtained by applying~\eqref{eq:rinvK} followed by the property (P3) of $GD$ as discussed in Sec.~\ref{Sec:GD} in the next step. Thus, the $GD$ between the observation and a roll-invariant target is a roll-invariant quantity. 

Table~\ref{tab:K_models} presents the Kennaugh matrices for the elementary scatterers such as the trihedral, cylinder, dipole, dihedral, narrow dihedral, $\pm 1/4$-wave, left helix and right helix in the HV basis.

\begin{table}[hbt]
    \centering
    \caption{Kennaugh Matrices for Elementary Targets}
    \label{tab:K_models}
    \resizebox{0.8\columnwidth}{!}{%
        \begin{tabular}{lcccc}
            \toprule
            Target & Row 1 & Row 2 & Row 3 & Row 4\\ \midrule
            $\mathbf{K}_d$ & 1 0 0 0 & 0 1 0 0 & 0 0 -1 0 & 0 0 0 1\\ 
            $\mathbf{K}_{nd}$ & 5/8 3/8 0 0 & 3/8 5/8 0 0 & 0 0 -1/2 0 & 0 0 0 1/2\\ 
            $\mathbf{K}_{t}$ & 1 0 0 0 & 0 1 0 0 & 0 0 1 0 & 0 0 0 -1\\ 
            $\mathbf{K}_{c}$ & 5/8 3/8 0 0 & 3/8 5/8 0 0 & 0 0 1/2 0 & 0 0 0 -1/2\\
            $\mathbf{K}_{dp}$ & 1 -1 0 0 & -1 1 0 0 & 0 0 0 0 & 0 0 0 0\\ 
            $\mathbf{K}_{+1/4}$ & 1 0 0 0 & 0 1 0 0 & 0 0 0 1 & 0 0 1 0\\ 
            $\mathbf{K}_{-1/4}$ & 1 0 0 0 & 0 1 0 0 & 0 0 0 -1 & 0 0 -1 0\\ \midrule
            $\mathbf{K}_{lh}$ & 1 0 0 -1 & 0 0 0 0 & 0 0 0 0 & -1 0 0 1\\ 
            $\mathbf{K}_{rh}$ & 1 0 0 1 & 0 0 0 0 & 0 0 0 0 & 1 0 0 1\\
            \bottomrule
        \end{tabular}
    }
\end{table}        
Among these elementary scattering models, the trihedral and helices are roll-invariant. 
\subsection{Alpha angle $\alpha_{GD}$}

In PolSAR literature, the scattering type angle $\alpha$ given by Cloude and Pottier is a roll invariant quantity~\cite{Cloude97} which varies in \SIrange{0}{90}{\degree}, corresponding to trihedral and dihedral scattering in the extremities, respectively. 
Alternately, the scattering type can be interpreted as the deviation/dissimilarity w.r.t. trihedral scattering. 
Using this interpretation, we define a new parameter $\alpha_{GD}$:
\begin{equation}
\alpha_{GD}(\mathbf{K}) = \SI{90}{\degree} \times GD(\mathbf{K},\mathbf{K}_t).
\end{equation}
The multiplication by \SI{90}{\degree} equals the scale for comparison with $\alpha$. 
For dihedral target, $\alpha_{GD}(\mathbf{K}_d) = \SI{90}{\degree}$ which matches with the scattering at extremities of $\alpha$ scale. 
Table~\ref{tab:alpha_tau_angles} shows $\alpha_{GD}$ for other elementary targets. 

Three clusters of elementary target models on the $\alpha_{GD}$ scale are evident. 
They are trihedral and cylinder; dipole and quarter wave devices; and lastly, narrow dihedral, dihedral, and helices.
        \begin{table}
            \centering
            \caption{$\alpha_{GD}$ and $\tau_{GD}$ for elementary targets}
            \begin{tabular}{lcc}\toprule
                Target & ${\alpha}_{GD}$ $[\SI{0}{\degree} , \SI{90}{\degree}]$ & $\tau_{GD}$ $[\SI{0}{\degree} , \SI{45}{\degree}]$\\ \midrule
                $\mathbf{K}_{rh}$ & $90$ & $45$\\
                $\mathbf{K}_{lh}$ &  $90$ & $45$\\ 
                $\mathbf{K}_d$ &  $90$ & $15$\\ 
                $\mathbf{K}_{nd}$ & $84.26$ & $13.37$\\ 
                $\mathbf{K}_{+1/4}$ &  $60$ & $7.24$\\ 
                $\mathbf{K}_{-1/4}$ & $60$ & $7.24$\\ 
                $\mathbf{K}_{dp}$ & $60$ & $7.24$\\
                $\mathbf{K}_{c}$ &  $25.84$ & $1.43$\\
                $\mathbf{K}_{t}$ & $0$ & $0$\\ \bottomrule  
            \end{tabular}
            \label{tab:alpha_tau_angles}
        \end{table}

\subsection{Helicity $\tau_{GD}$}
The helicity parameter provides a quantitative estimation of target symmetry~\cite{Touzi2007} in the observation. 
This quantity is derived from $f_{lh}$ and $f_{rh}$, i.e., the individual similarity of the observation with left and right helix models respectively. 
A single measure of helicity is obtained by replacing $GD$ with the (geometric) mean of the distances from left and right helices in the definition of similarity:
\begin{equation}
\tau_{GD} = \SI{45}{\degree} \times \left(1 - \sqrt{GD(\mathbf{K},\mathbf{K}_{lh})GD(\mathbf{K},\mathbf{K}_{rh})}\right).
\end{equation}
The multiplication by \SI{45}{\degree} makes the scale equal for comparison with $|\tau_{m_1}|$~\cite{Touzi2007}. 
For trihedral target, $\tau_{GD} = 0$ and for helices $\tau_{GD} = \SI{45}{\degree}$, which matches with the scattering at extremities of $|\tau_{m_1}|$. 
Table~\ref{tab:alpha_tau_angles} provides the helicity values for other elementary targets. 
It is observed that $\tau_{GD}$ discriminates between helices and dihedral, in addition to the discrimination of elementary targets as provided by $\alpha_{GD}$.

\subsection{Purity Index $P_{GD}$}
The ensemble averaging of the Stokes matrix was explored in~\cite{Fry81}. 
Under such circumstances~\eqref{FWE} turns into an inequality:
\begin{equation}\label{FWIE}
4K_{11}^2  \geq \sum_{j=1}^{4} K_{ij}^2 .
\end{equation}
Later, the Stokes-Mueller formalism was revisited by Barakat~\cite{barakat81} and Simon~\cite{simon82} to obtain equivalent equations in a different mathematical setting for a fully polarized system. 
Using~\eqref{FWE} and its physical significance as given in~\cite{simon82}, Gil and Bernabeau identified its potential as a criterion for depolarization using Mueller matrices~\cite{gil85}. 
They defined a depolarization index (presented here using the Kennaugh matrix) as:
\begin{equation}\label{gildep}
P_D = \sqrt{\frac{\Tr(\mathbf{K}^T\mathbf{K}) - K_{11}^2}{3K_{11}^2}} ,
\end{equation} 
where $P_D$ being $1$ and $0$ corresponds to a non-depolarizing media and the ideal depolarizer, respectively. 
The Kennaugh matrix form for the ideal depolarizer~\cite{cloude2010polarisation} is:
\begin{equation}
\mathbf{K}_{dep} = \left[\begin{array}{cccc}
1 & 0 & 0 & 0\\
0 & 0 & 0 & 0\\
0 & 0 & 0 & 0\\
0 & 0 & 0 & 0
\end{array}\right].
\end{equation}
It can be shown that there exists no corresponding $\mathbf{S}_{dep}$ matrix for $\mathbf{K}_{dep}$. Hence, a coherent physical target does not exist for the ideal depolarizer in nature. 
Eq.~\eqref{gildep} can also be rewritten as:
\begin{equation}
P_D = \frac{1}{\sqrt{3}} \norm{\left( \frac{\mathbf{K}}{K_{11}} - \mathbf{K}_{dep} \right) },
\end{equation}
where $\|\cdot\|$ denotes the Euclidean norm of a real matrix. 
Thus, $P_D$ is obtained by measuring the distance between $K_{11} = \text{Span}/2$ (normalized) observation and the ideal depolarizer. 
It may be noted that $\mathbf{K}_{dep}$ is invariant under the roll transformation. 
Thus, it is important to investigate the limits of $GD(\mathbf{K},\mathbf{K}_{dep})$ by simplifying the expression:
$$
\frac{\Tr(\mathbf{K}^T\mathbf{K_{dep}})}{\sqrt{\Tr(\mathbf{K}^T\mathbf{K})\Tr(\mathbf{K_{dep}^T}\mathbf{K_{dep}})}} =  \frac{K_{11}}{\sqrt{\Tr(\mathbf{K}^T\mathbf{K})}}.
$$
Using~\eqref{FWIE} we obtain that
$$
\frac{K_{11}}{\sqrt{\Tr(\mathbf{K}^T\mathbf{K})}} \geq \frac{1}{2}.
$$
This implies that
$$
GD(\mathbf{K},\mathbf{K}_{dep}) \leq  \frac{2}{\pi}\cos^{-1}\left(\frac{1}{2}\right) = \frac{2}{3}.
$$
Thus, $GD(\mathbf{K},\mathbf{K}_{dep})$ varies in the range $[0,{2}/{3}]$ with zero corresponding to the ideal depolarizer and ${2}/{3}$ corresponding to non-depolarizing media. 
Thus the resulting quantity $\frac{3}{2}GD(\mathbf{K},\mathbf{K}_{dep})$ is the dissimilarity between the observation and $\mathbf{K}_{dep}$. 
This quantity is large even for distributed targets. 
Hence, to have a good contrast we take the square of this quantity as our definition for the depolarization index:
\begin{equation}
P_{GD} = \left(\frac{3}{2}GD(\mathbf{K},\mathbf{K}_{dep})\right)^2,
\label{depi}
\end{equation}
where $P_{GD} = 0$ corresponds to the ideal depolarizer and $P_{GD} = 1$ corresponds to non-depolarizing targets. 
All coherent targets shown in Table~\ref{tab:K_models} have $P_{GD} = 1$.   

These three roll-invariant parameters along with the $\text{Span}$ can be utilized to classify a PolSAR scene and interpret the scattering type. 

A disadvantage with roll-invariant parameters is their inability to separate the quarter waves from the dipole. 
In the next section, we compare the proposed parameters with some well-known parameters from the PolSAR literature.             

\subsection{Comparisons with parameters from literature}

We have derived the three roll invariant parameters $\alpha_{GD}$, $\tau_{GD}$ and $P_{GD}$ using the geodesic distance. Although these parameters are obtained from a different formulation their interpretation is similar to specific well-known parameters in PolSAR literature~\cite{Touzi2007,Cloude97,gil85}.

\begin{figure}[hbt]
    \centering
    \subfloat[$\alpha_{GD}$]{\includegraphics[width=0.48\columnwidth]{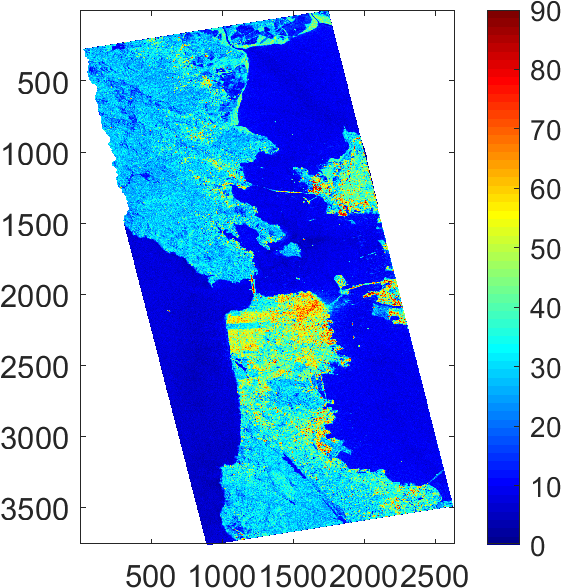}\label{subfig:alp_gd_SF}}~
    \subfloat[$\alpha$]{\includegraphics[width=0.48\columnwidth]{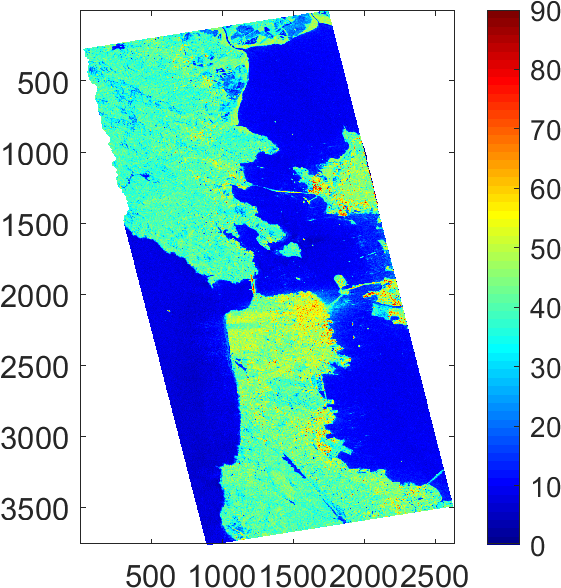}\label{subfig:alp_SF}}\\
    
    \subfloat[$\tau_{GD}$]{\includegraphics[width=0.48\columnwidth]{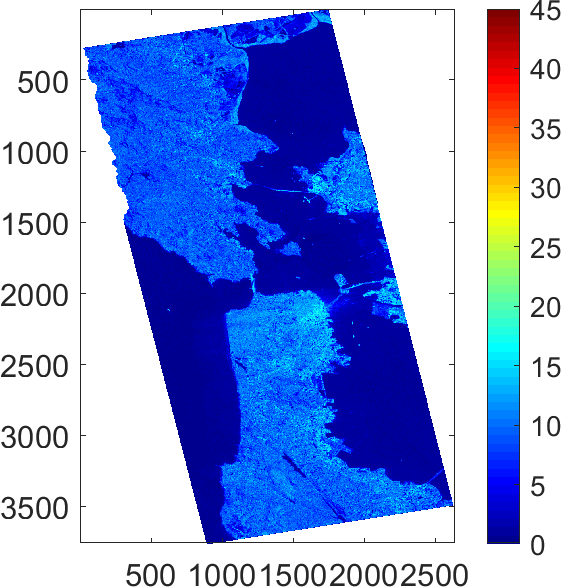}\label{subfig:hel_gd_SF_RS2}}~
    \subfloat[$|\tau_{m_1}|$]{\includegraphics[width=0.48\columnwidth]{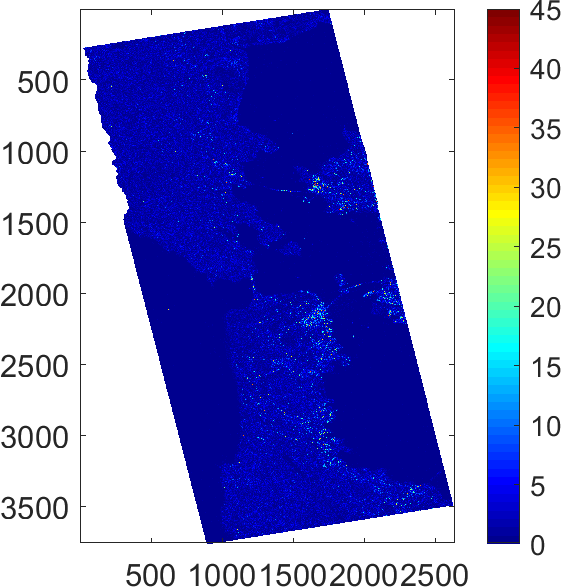}\label{subfig:hel_SF_RS2}}\\
    
    \subfloat[$P_{GD}$]{\includegraphics[width=0.48\columnwidth]{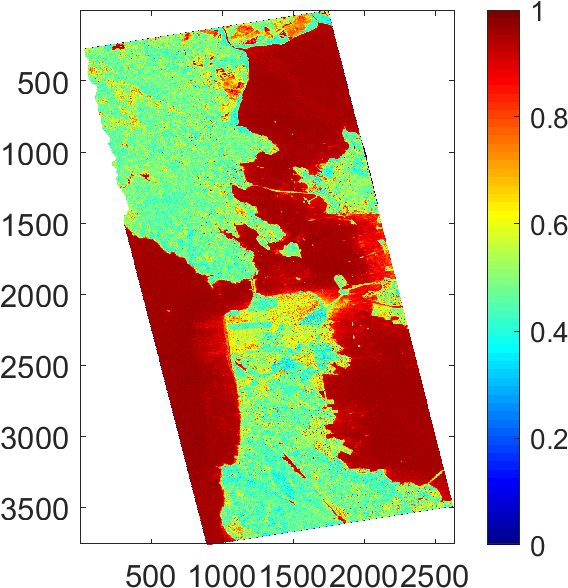}\label{subfig:p_gd_SF_RS2}}~
    \subfloat[$P_{D}$]{\includegraphics[width=0.48\columnwidth]{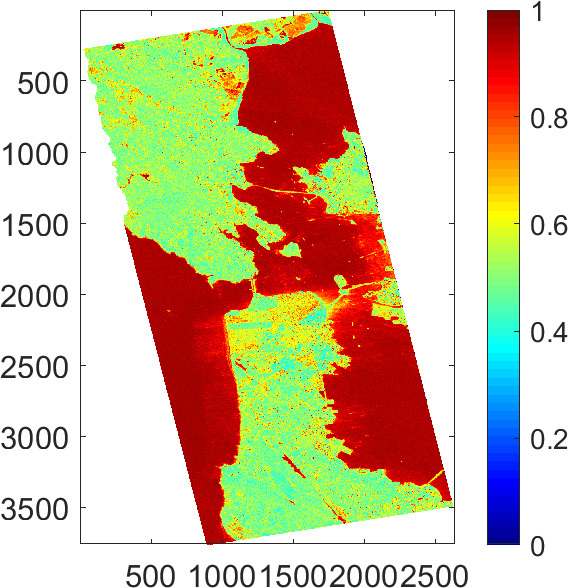}\label{subfig:p_d_SF_RS2}}\\
    \caption{Parameter values for proposed and existing parameters for RS-2 C-Band SF Image}
    \label{fig:RS2_par_maps}
\end{figure}

\begin{figure}[hbt]
    \centering
    
    \subfloat[$\alpha_{GD}$]{\includegraphics[width=0.48\columnwidth]{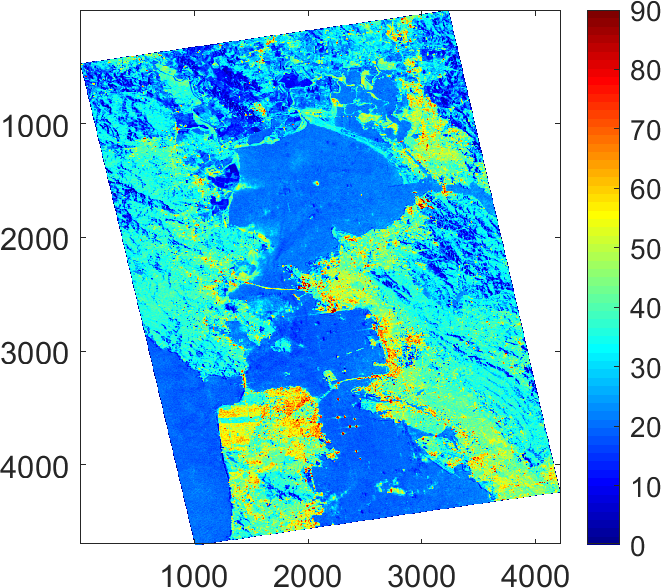}\label{subfig:alp_gd_SF_ALOS2}}~
    \subfloat[$\alpha$]{\includegraphics[width=0.48\columnwidth]{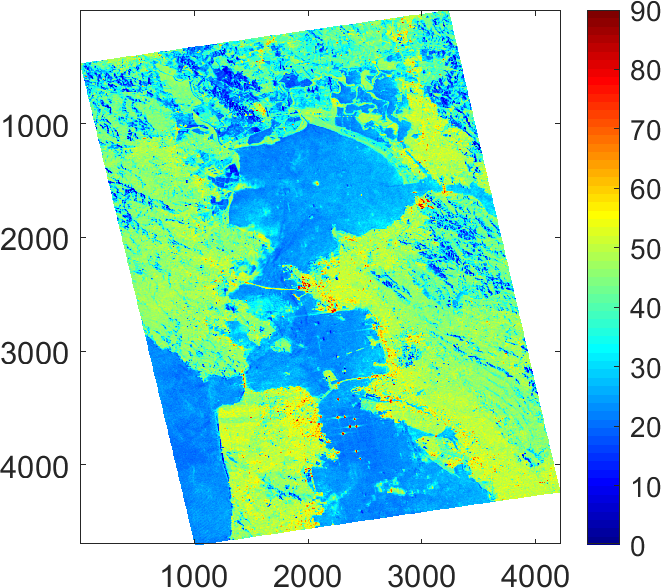}\label{subfig:alp_SF_ALOS2}}\\
    
    \subfloat[$\tau_{GD}$]{\includegraphics[width=0.48\columnwidth]{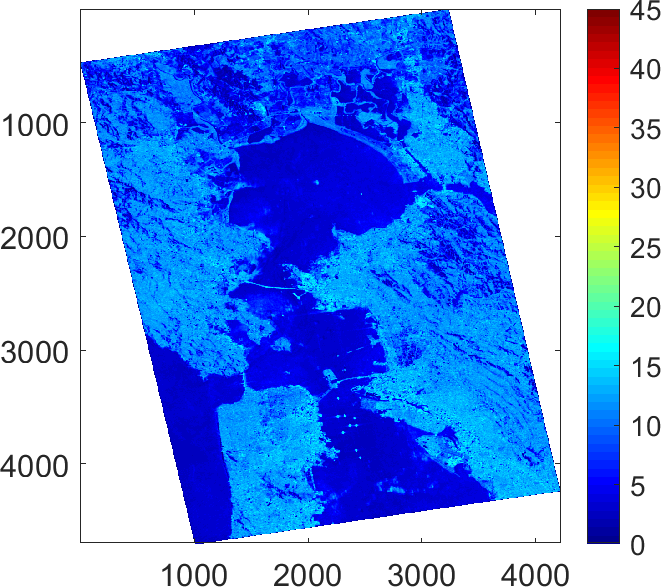}\label{subfig:hel_gd_SF_ALOS2}}~
    \subfloat[$|\tau_{m_1}|$]{\includegraphics[width=0.48\columnwidth]{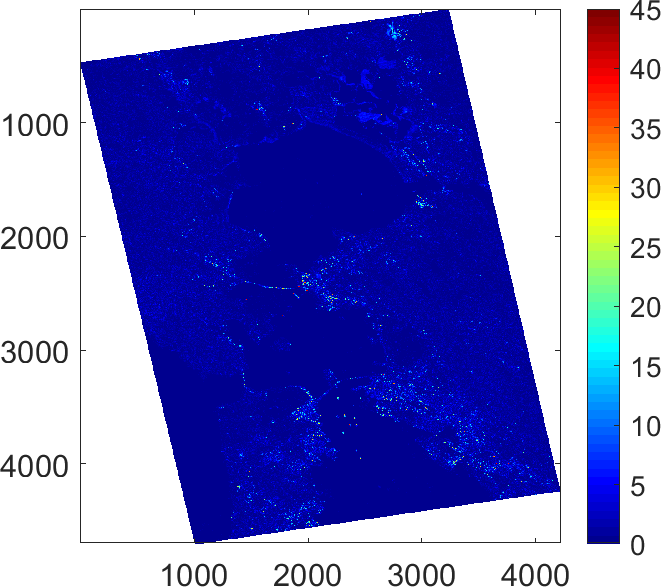}\label{subfig:hel_SF_ALOS2}}\\
    
    \subfloat[$P_{GD}$]{\includegraphics[width=0.48\columnwidth]{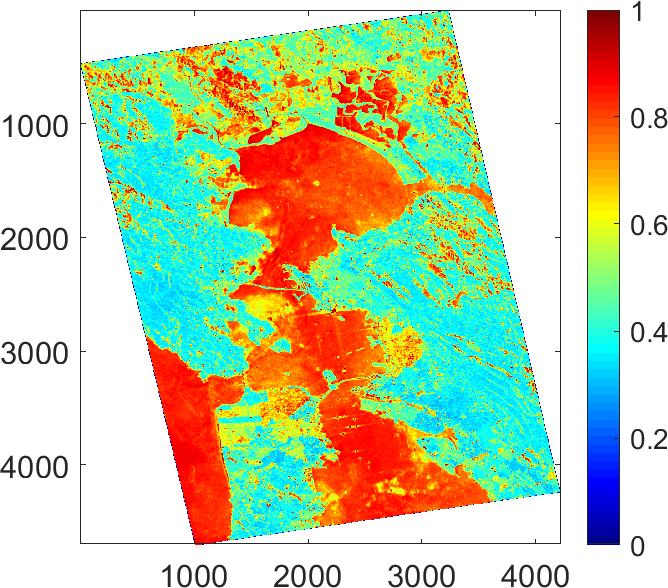}\label{subfig:p_gd_SF_ALOS2}}~
    \subfloat[$P_{D}$]{\includegraphics[width=0.48\columnwidth]{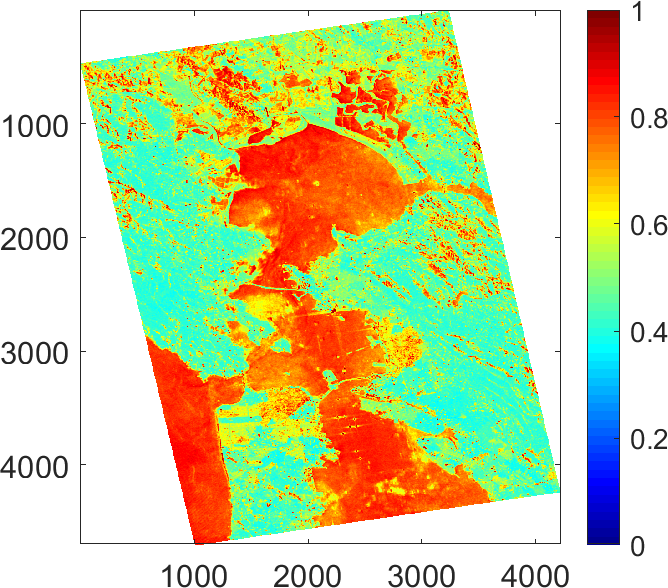}\label{subfig:p_d_SF_ALOS2}}\\
    
    \caption{Parameter values for proposed and existing parameters for ALOS-2 L-band SF Image}
    \label{fig:ALOS2_par_maps}
\end{figure}

Figs.~\ref{fig:RS2_par_maps} and~\ref{fig:ALOS2_par_maps} respectively show the proposed parameters and the corresponding parameters from PolSAR literature for RS-2 C-band and ALOS-2 L-band images of San Francisco in pairs. 
It is observed that $\alpha_{GD}$ has a more dynamic range in comparison to $\alpha$, leading to better discrimination of different scatterers. 
The $\tau_{GD}$ has a higher value than $|\tau_{m_1}|$, especially pronounced over land masses. 
The purity parameters $P_{GD}$ and $P_D$ look identical.

\begin{figure}[hbt]
    
    \subfloat[Ref. row of RS-2 image]{\includegraphics[width=0.48\columnwidth]{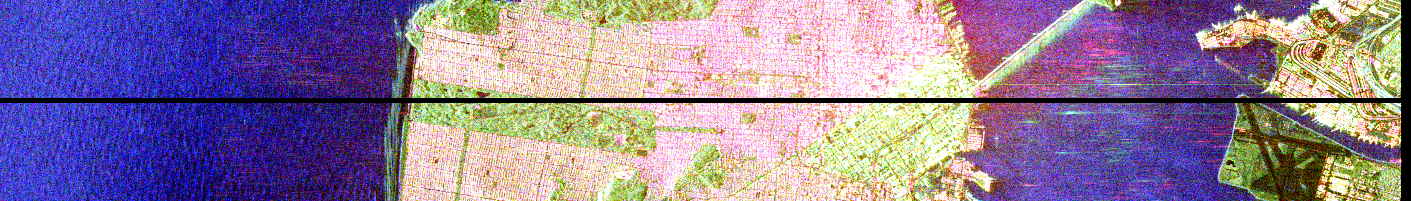}\label{subfig:row_SF_RS2}}\hspace{0.25em}               
    \subfloat[Ref. row of ALOS-2 image]{\includegraphics[width=0.48\columnwidth]{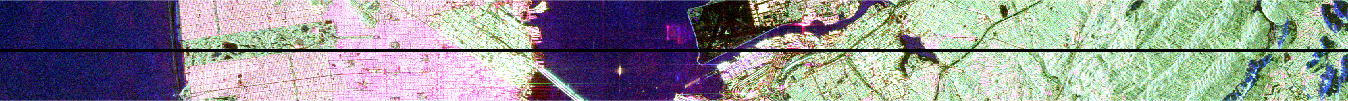}\label{subfig:row_SF_ALOS2}}\\

    \subfloat[$\alpha_{GD}$ vs $\alpha$]{\includegraphics[width=0.48\columnwidth]{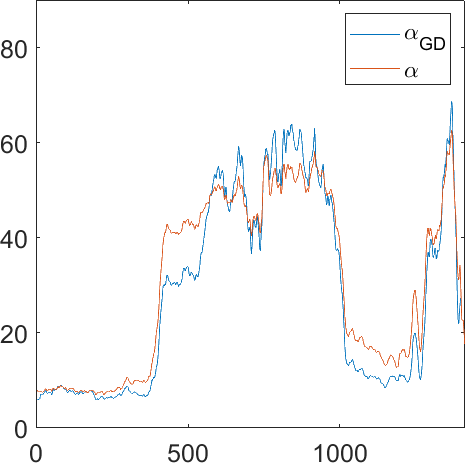}\label{subfig:alp_comp}}\hspace{0.25em}
    \subfloat[$\alpha_{GD}$ vs $\alpha$]{\includegraphics[width=0.48\columnwidth]{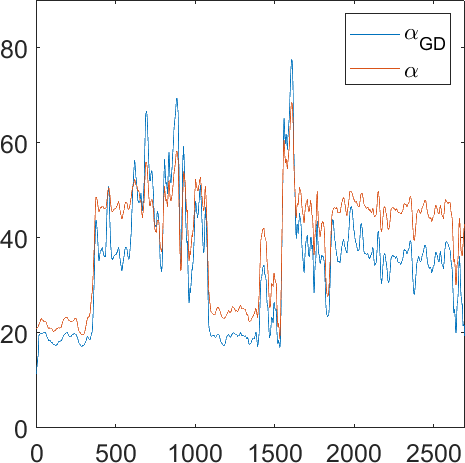}\label{subfig:alp_comp_ALOS2}}\\
    
    \subfloat[$\tau_{GD}$ vs $|\tau_{m1}|$]{\includegraphics[width=0.48\columnwidth]{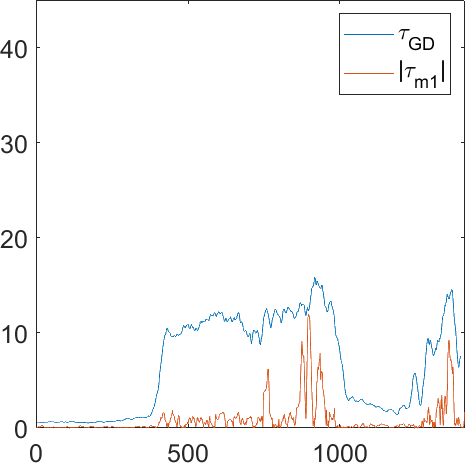}\label{subfig:hel_comp}}\hspace{0.25em}
    \subfloat[$\tau_{GD}$ vs $|\tau_{m1}|$]{\includegraphics[width=0.48\columnwidth]{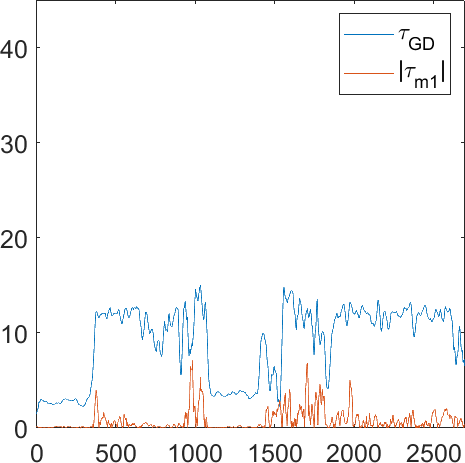}\label{subfig:hel_comp_ALOS2}}\\
    
    \subfloat[$P_{GD}$ vs $P_D$]{\includegraphics[width=0.48\columnwidth]{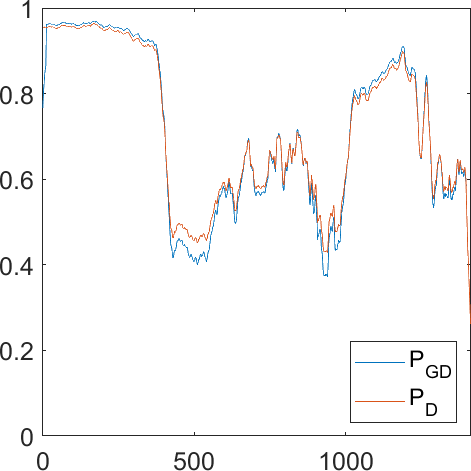}\label{subfig:depi_comp}}\hspace{0.25em}
    \subfloat[$P_{GD}$ vs $P_D$]{\includegraphics[width=0.48\columnwidth]{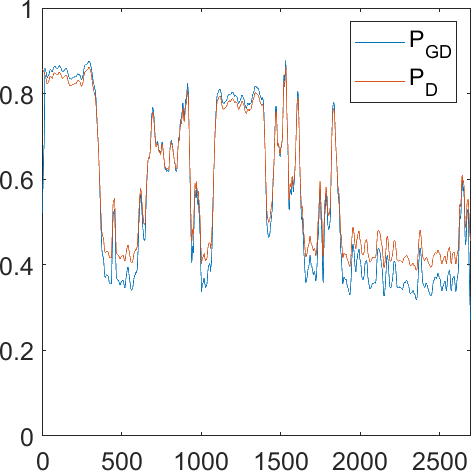}\label{subfig:depi_comp_ALOS2}}\\
    
    \caption{Profile of roll-invariant parameters computed along particular transects in the RS-2 and the ALOS-2 SF images}
    \label{fig:Transact_Var}
\end{figure}

For a more in-depth understanding, we performed a quantitative assessment of the above parameters in Fig.~\ref{fig:Transact_Var}. 
The figure shows the transect for two particular rows for RS-2 and ALOS-2 data sets respectively. 
The transect passes through the Golden Gate Park of San Francisco and the South Market Area, i.e., the oriented urban buildings elusive to most target decompositions under identification. 
The transect contains several scattering zones: sea-surface, vegetation, urban area block perpendicular to radar LoS and those oriented to it.

It can be seen from the transects that the $\alpha_{GD}$ has a better dynamic range than $\alpha$. 
It is lower for scattering from sea surface (\SIrange{0}{25}{\degree}) and vegetation (\SIrange{20}{40}{\degree}), and higher for urban areas ($>\SI{40}{\degree}$) than $\alpha$. 
The value of $\tau_{GD}$ is higher than $\tau$ throughout the transect for all the types scatters. 
However, a marked jump is seen over land surfaces where $\tau_{GD} > \SI{10}{\degree}$. 
It is also observed that $\tau_{GD}$ and $|\tau_{m_1}|$ are different to a great extent because of their respective definitions. 
However, both quantities are measures of the asymmetric nature of the scattering in the pixel, which makes them similar for comparison. 
The purity indices $P_{GD}$ and $P_{D}$ are very similar; however, $P_{GD}$ is slightly better than $P_D$ from purer to distributed targets.

\section{Scattering zone identification using Roll-Invariant Parameters}\label{Sec:Zone}

\begin{figure}[ht!b]
    \centering
    \subfloat[Histogram of  $\alpha_{GD}$ for RS-2 image]{\includegraphics[width=\columnwidth]{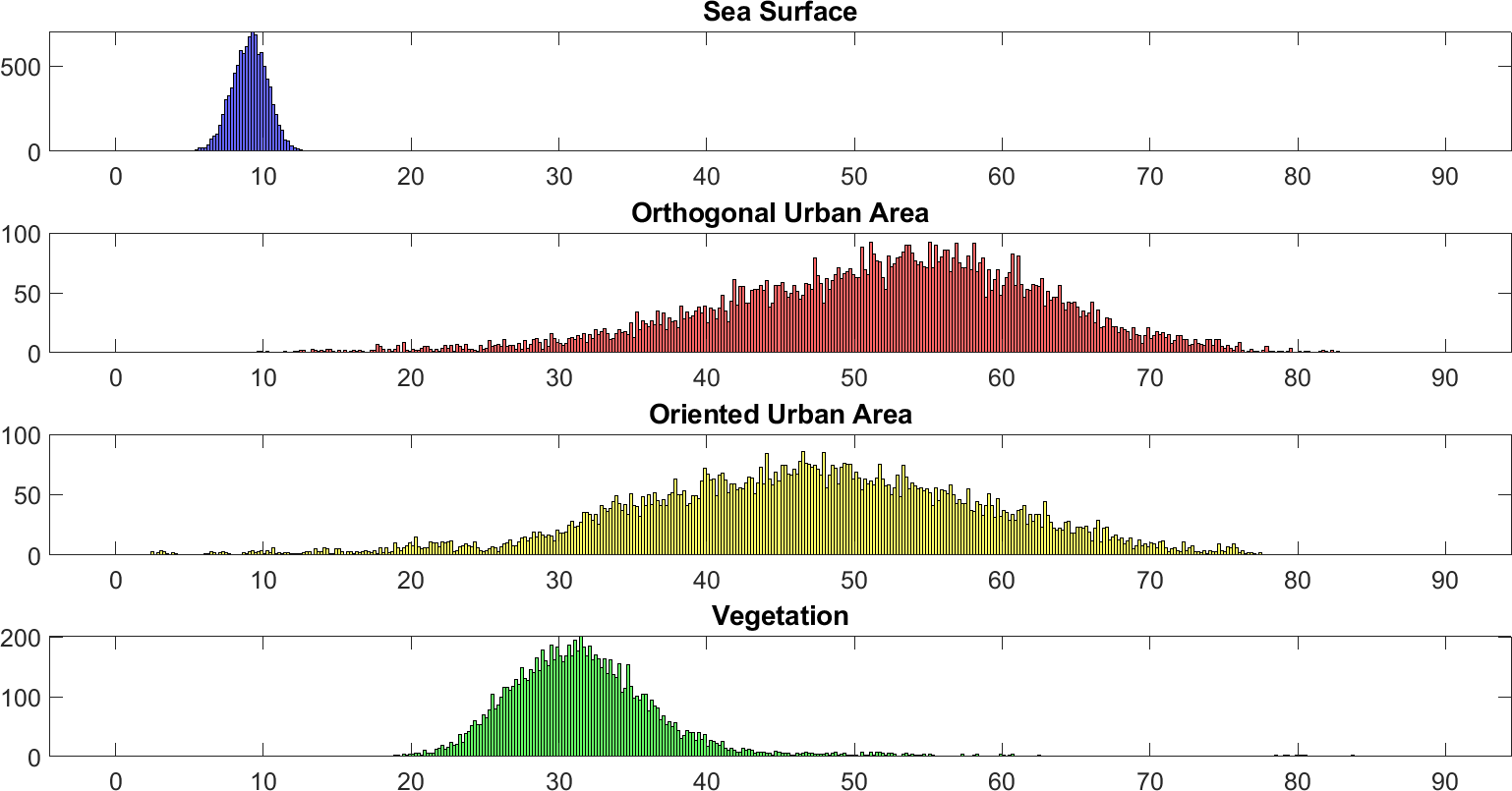}\label{fig: Hist_RS2_a}}\\
    \subfloat[Histogram of  $\alpha_{GD}$ for ALOS-2 image]{\includegraphics[width=\columnwidth]{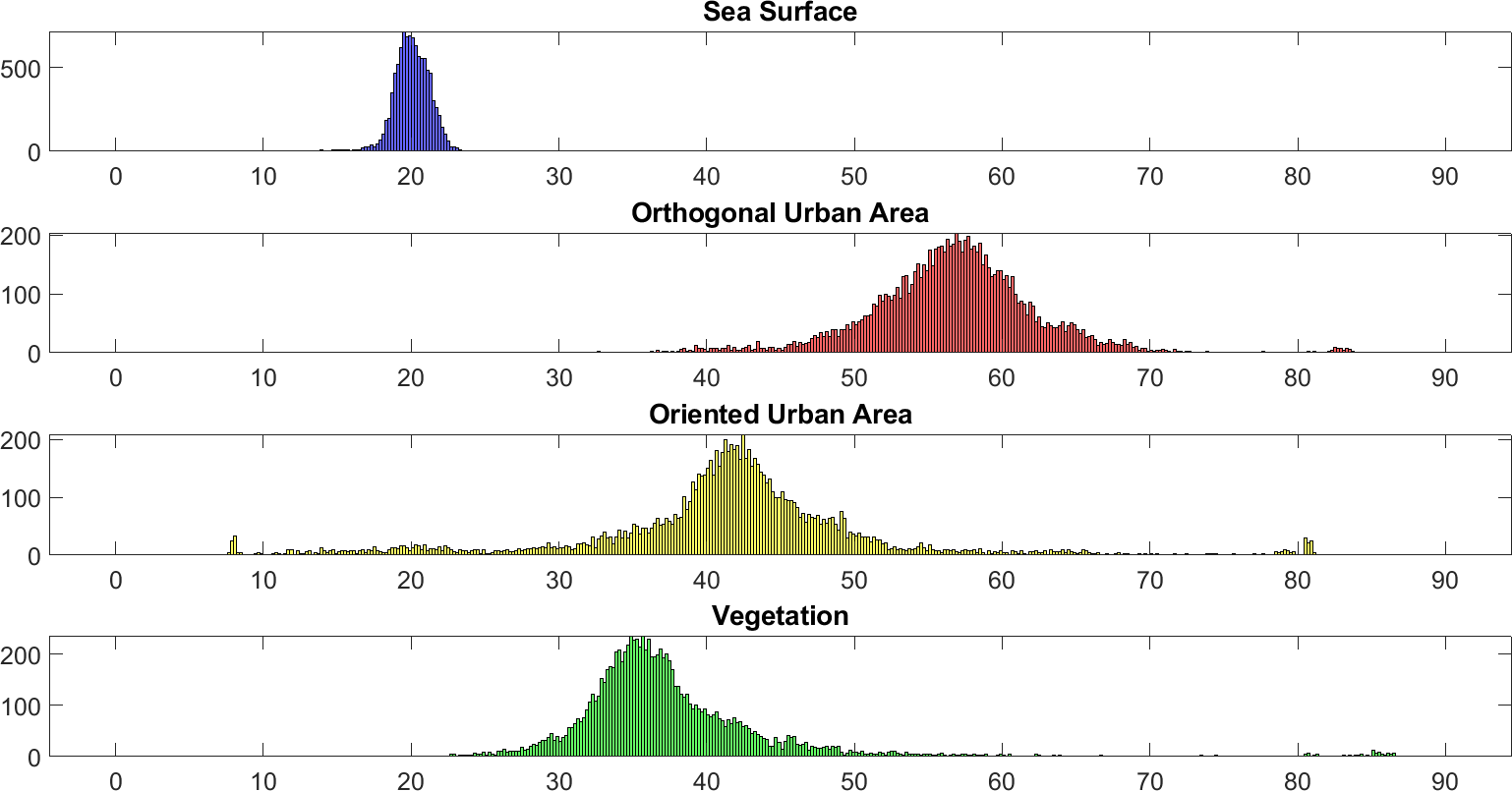}\label{fig: Hist_ALOS2_a}}\\
    \subfloat[Histogram of  $\tau_{GD}$ for RS-2 image]{\includegraphics[width=\columnwidth]{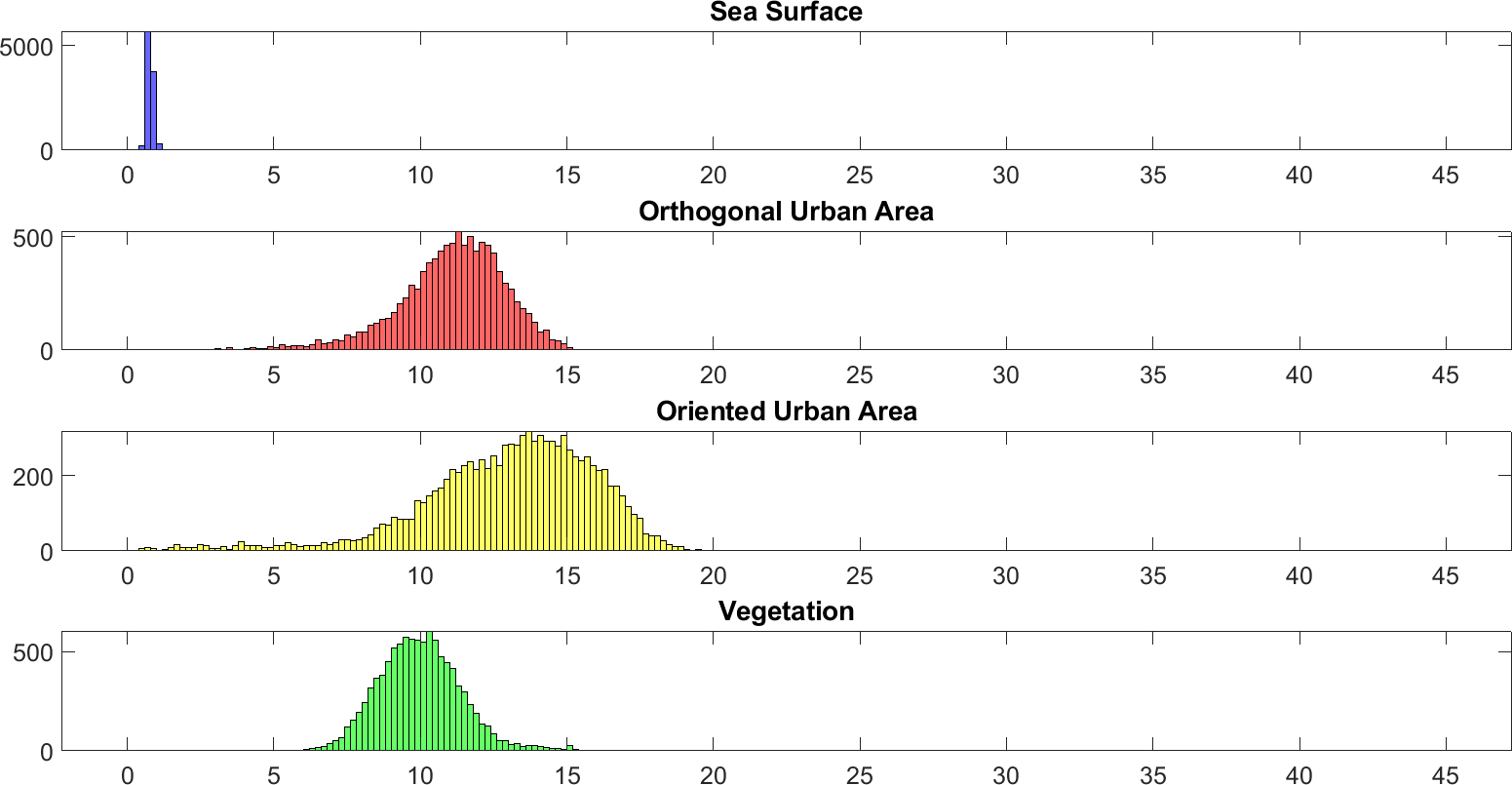}\label{fig: Hist_RS2_t}}\\
    \subfloat[Histogram of  $\tau_{GD}$ for ALOS-2 image]{\includegraphics[width=\columnwidth]{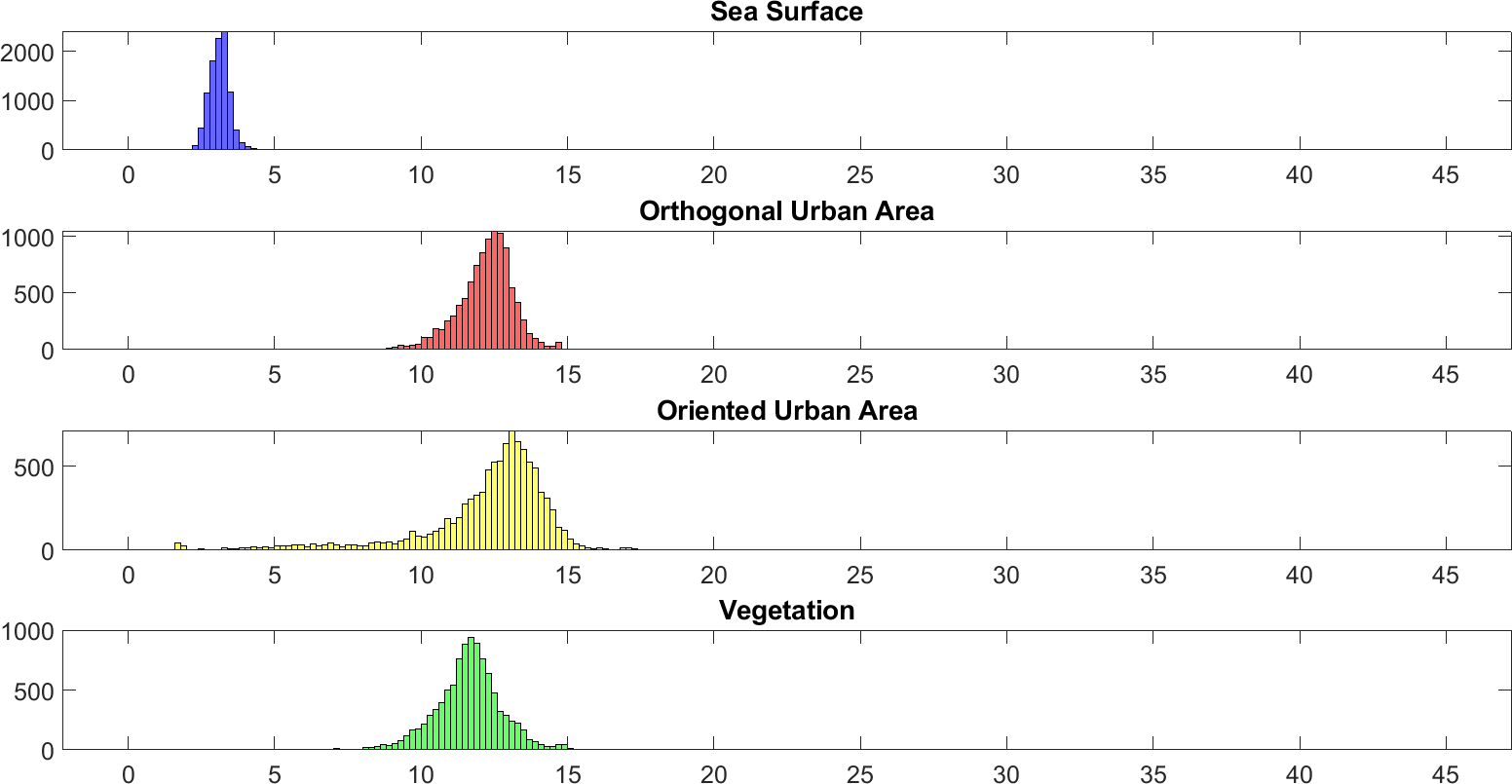}\label{fig: Hist_ALOS2_t}}\\
    \caption{Histograms for $\alpha_{GD}$ and $\tau_{GD}$ for different scattering zones for RS-2 and ALOS-2 SF images.}
    \label{fig:Hist_pars}
\end{figure}
\begin{figure}[ht!]
    \centering
    
    ~\subfloat[Segmentation with $\alpha_{GD}$]{\includegraphics[width=0.415\columnwidth]{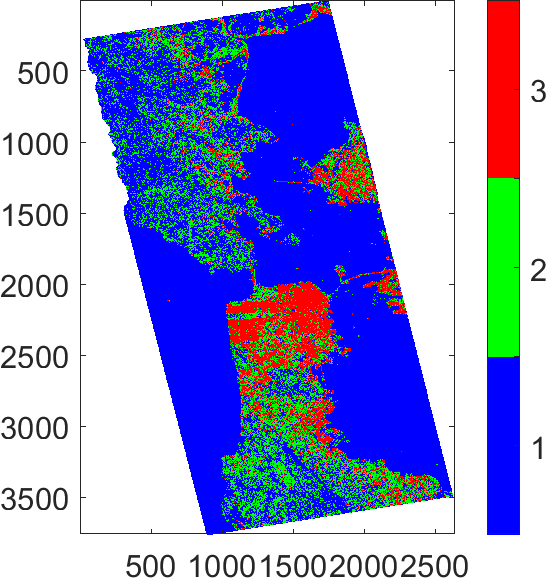}\label{fig:alp_gd_seg_sf_rs2}}
    ~\subfloat[Segmentation with $\alpha_{GD}$]{\includegraphics[width=0.50\columnwidth]{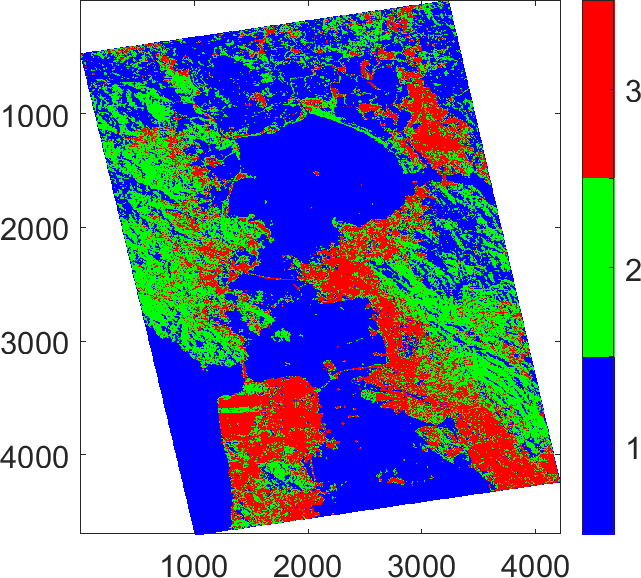}\label{fig:alp_gd_seg_sf_alos2}}\\                
    
    ~\subfloat[Segmentation with $\tau_{GD}$]{\includegraphics[width=0.415\columnwidth]{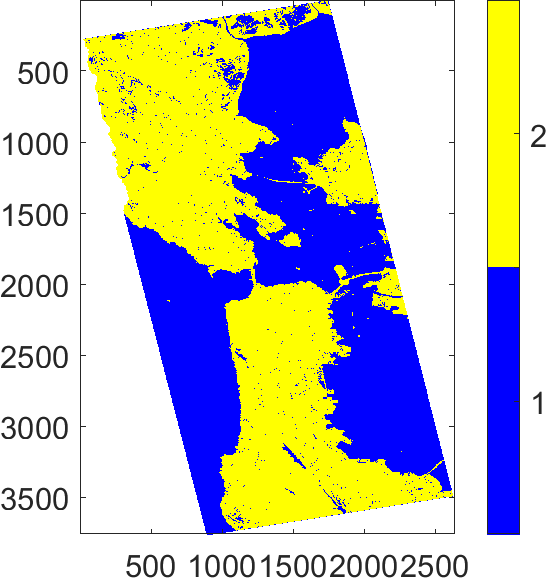}\label{fig:tau_gd_seg_sf}}
    ~\subfloat[Segmentation with $\tau_{GD}$]{\includegraphics[width=0.50\columnwidth]{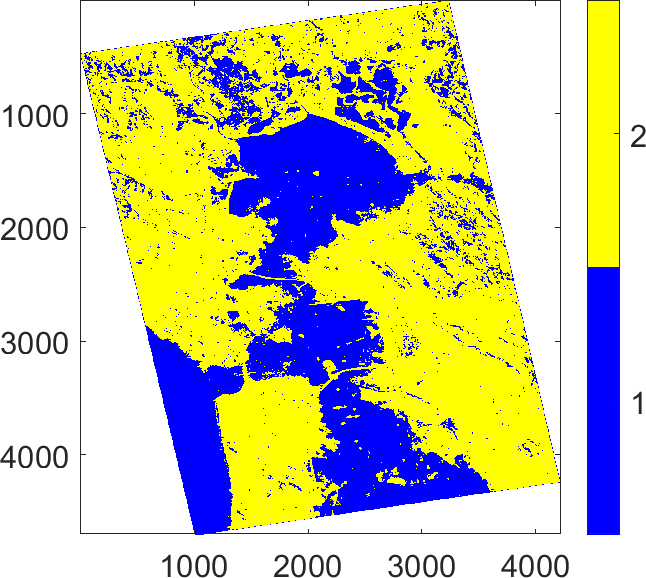}\label{fig:tau_gd_seg_mm}}\\

    \caption{Segmentation results}
    \label{fig:Seg_inv_par_SF}
\end{figure}

In this section, we utilize the roll-invariant parameters for unsupervised classification of the PolSAR scene. At first, we assess the nature of classification provided by the parameters~$\alpha_{GD}$ and $\tau_{GD}$ independently. And finally we propose a $P_{GD}/\alpha_{GD}$ scheme for unsupervised classification for PolSAR images using both these parameters simultaneously. 

We begin by recalling Table~\ref{tab:alpha_tau_angles} which shows the values of $\alpha_{GD}$ and $\tau_{GD}$ for various elementary scatterers. 
Among them, the family of coherent odd-bounce scatters is restricted to within \SI{30}{\degree} on the $\alpha_{GD}$ scale and less than \SI{5}{\degree} on the $\tau_{GD}$ scale. 
This kind of scattering is mainly seen from the sea surface. 
For the rest of the elementary scatterers except helices, the $\tau_{GD}$ is restricted to the closed range \SIrange{5}{15}{\degree}. 
Thus, segmenting $\tau_{GD}$ into two parts, namely below and above \SI{5}{\degree}, separates the sea from the terrain. 

Another point of interest is the vegetation, which is a distributed target. 
In this respect, the Yamaguchi 3-case volume model~\cite{Yamaguchi2005} is a simple and popular model often used in PolSAR literature. 
The $\alpha_{GD}$ for the three cases are \SI{40.40}{\degree}, \SI{35.26}{\degree} and \SI{40.40}{\degree} respectively.
For the identification of urban areas, the even-bounce family of narrow dihedral and dihedral are often employed.
Ref.~\cite{Yamaguchi2005} introduces the helix component into the decomposition to account for the co-pol and cross-pol correlation in the urban areas. 
Hence, on the $\alpha_{GD}$ scale the odd-bounce scatterers occur first, then the volume scatterers and finally the even bounce and helix scatterers. 
Based on this analysis, we have good estimates of the segments that the $\alpha_{GD}$ scale can be broken into for identification of these scattering zones: 
$[\SI{0}{\degree}, \SI{30}{\degree})$, 
$[\SI{30}{\degree}, \SI{40}{\degree})$, 
and
$[\SI{40}{\degree}, \SI{90}{\degree}]$.  
         
We further validate the choice of these segments by plotting the histograms for $\alpha_{GD}$ and $\tau_{GD}$ for samples from different scattering zones within the PolSAR images in Fig.~\ref{fig:Hist_pars}. 
These samples are representative areas of the sea surface, orthogonal urban area, oriented urban area and, vegetation.

The largest overlap between $\alpha_{GD}$ histograms occurs for vegetation and oriented urban areas. 
Resolving this ambiguity is a significant avenue of research within PolSAR literature~\cite{Thirion2017XPol,Chen2014}. 
The peaks of the distribution for different scattering zones lie in the exclusive segments that we hypothesized for $\alpha_{GD}$. 
The segmentation with \SI{5}{\degree} is an effective way of segmenting a PolSAR image by using $\tau_{GD}$ because of the wide separation between the odd-bounce scatterers and other scatterers on its scale.

Figure~\ref{fig:Seg_inv_par_SF} shows the classification achieved using the above segments for $\alpha_{GD}$ and $\tau_{GD}$ individually. 
The $\alpha_{GD}$ is able to extract the oriented urban areas from the San Francisco scene in both the data sets, while $\tau_{GD}$ identifies the land and ships from the sea accurately.

\section{$P_{GD}/\alpha_{GD}$ Classification scheme}\label{Sec:Classification}

We have verified the usefulness of $\alpha_{GD}$ for urban mapping. 
It will be beneficial if this segmentation of $\alpha_{GD}$ is combined with other roll-invariant parameters to form a more detailed classification scheme. 

We chose $P_{GD}$ as the second parameter. 
A similar kind of classification scheme was first attempted in~\cite{Praks2009}, to provide an alternative to the original $H/\alpha$ classification~\cite{Cloude97}. 
The parameters utilized were the surface scattering fraction and the scattering diversity, the latter being related to the degree of purity $P_{D}$. 
The authors utilized ranges following the $H/\alpha$ scheme, a choice that led to suboptimal exploitation of the potential of their work.
In this light, we stipulated the parameters' segments according to their behavior over well-known scattering zones as obtained in our previous analysis.

We chose four segments of $\alpha_{GD}$ instead of three.
The last segment $[\SI{40}{\degree}, \SI{90}{\degree}]$ is divided in two: $[\SI{40}{\degree}, \SI{80}{\degree})$ and $[\SI{80}{\degree}, \SI{90}{\degree}]$.
Canonical scatterers like the narrow dihedral, dihedral, and helices concentrate in the last interval.
The $P_{GD}$ is split in the middle at $0.5$ to discriminate coherent targets from the rest. 

Fig.~\ref{fig:p_a_seg_SF} shows the scatterplot of $(P_{GD},\alpha_{GD})$ and the classification results for the RS-2 C-band and ALOS-2 L-band San Francisco images. 
The classes are based on the Euclidean product of segments of $\alpha_{GD}$ and $P_{GD}$ listed in Table~\ref{tab:Class_ref}.

\begin{table}[hbt]
	\centering
	\caption{$P_{GD}/\alpha_{GD}$ classes}
	\begin{tabular}{lc@{\hspace{.6em}}c@{\hspace{.6em}}c@{\hspace{.6em}}c}\toprule
		$ \alpha_{GD} \in$ & $ [\SI{0}{\degree}, \SI{30}{\degree})$ & 
		$[\SI{30}{\degree}, \SI{40}{\degree})$ & 
		$[\SI{40}{\degree}, \SI{80}{\degree})$ & 
		$[\SI{80}{\degree}, \SI{90}{\degree}]$\\\midrule
		$P_{GD} \leq 0.5$ & 1 & 3 & 5 & 7\\
		$P_{GD} > 0.5$ &  2 & 4 & 6 & 8\\
		\bottomrule
	\end{tabular}
	\label{tab:Class_ref}
\end{table}

\subsection{Shape of the Scatter Plot}
In a similar manner to the computation of the feasible region for the $H/\alpha$ scatter plot~\cite{Cloude97}, we compute the theoretical bounds for the physical scatterers in terms of $P_{GD}/\alpha_{GD}$. 

\begin{figure}[hbt]
        \centering
        \subfloat[$P_{GD}$--$\alpha_{GD}$ segmentation]{\includegraphics[width=0.40\columnwidth]{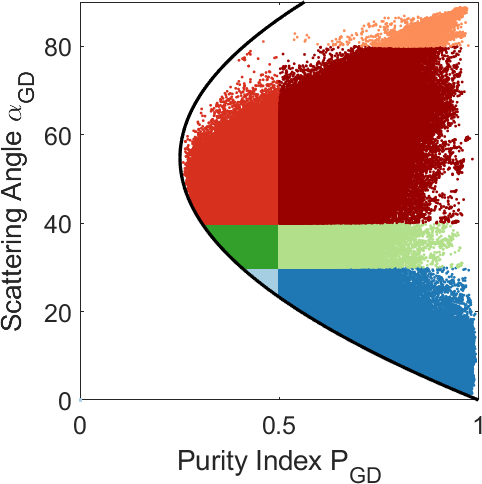}\label{fig:Alp_P_GD_plot_sf}}\qquad\quad
        \subfloat[$P_{GD}$--$\alpha_{GD}$ segmentation]{\includegraphics[width=0.40\columnwidth]{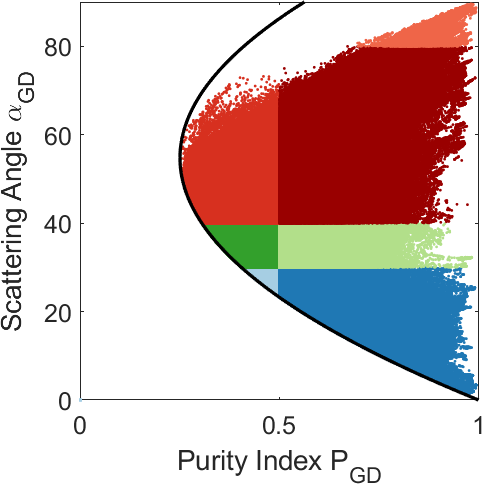}\label{fig:Alp_P_GD_plot_sf_ALOS2}}\\
        
        \subfloat[RS-2 segmentation map]{\includegraphics[width=0.415\columnwidth]{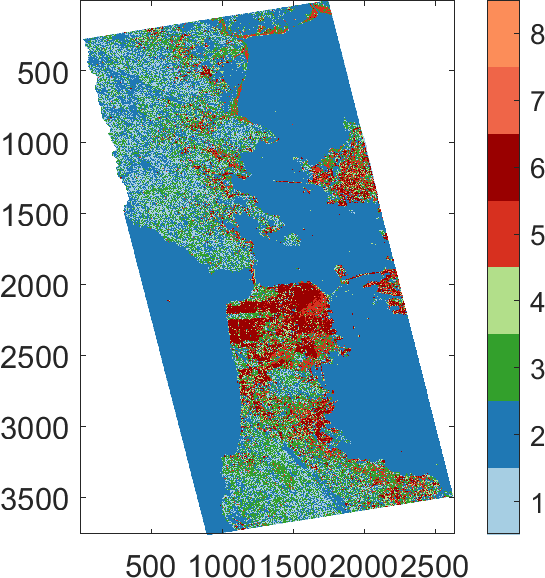}\label{fig:Alp_P_GD_Seg_sf}}\quad
        \subfloat[ALOS-2 segmentation map]{\includegraphics[width=0.50\columnwidth]{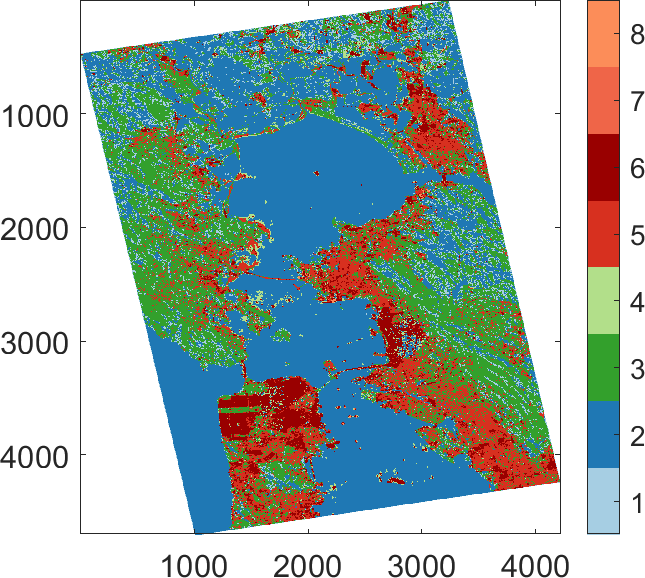}\label{fig:Alp_P_GD_Seg_sf_ALOS2}}
        \\                    
        
        \caption{Unsupervised classification using $P_{GD}$ and $\alpha_{GD}$}
        \label{fig:p_a_seg_SF}
    \end{figure}

For physical scatterers, the feasible region in the $P_{GD}/\alpha_{GD}$ plane is delimited by two curves namely I and II. The two curves are characterized by scatterers whose coherency matrices are given in~\cite{Cloude97}. We present their corresponding Kennaugh matrix forms using~\eqref{incoKen} as given below, 
\begin{eqnarray}
\mathbf{K}_{\text{I}} & = & \left[\begin{array}{cccc}
\frac{2m+1}{2} & 0 & 0\\
0 & \frac{1}{2} & 0 & 0\\
0 & 0 & \frac{1}{2} & 0\\
0 & 0 & 0 & \frac{2m-1}{2}\\
\end{array}\right] 0\leq m\leq 1,\nonumber\\
\mathbf{K}_{\text{II}} & = & \left[\begin{array}{cccc}
\frac{2m+1}{2} & 0 & 0\\
0 & \frac{1-2m}{2} & 0 & 0\\
0 & 0 & \frac{2m-1}{2} & 0\\
0 & 0 & 0 & \frac{2m+1}{2}\\
\end{array}\right] 0\leq m\leq 0.5,\nonumber\\
\mathbf{K}_{\text{III}} & = & \left[\begin{array}{cccc}
\frac{2m+1}{2} & 0 & 0\\
0 & \frac{2m-1}{2} & 0 & 0\\
0 & 0 & \frac{2m-1}{2} & 0\\
0 & 0 & 0 & \frac{3-2m}{2}\\
\end{array}\right] 0.5\leq m\leq 1.\nonumber\\
\end{eqnarray}
The curve I which bounds the scatter plot from below, in particular, called the azimuthal symmetry curve. We compute the $P_{GD}/\alpha_{GD}$ values for these scatterers and trace the curves (shown in black) within the scatter plot plane in the Fig.~\ref{fig:p_a_seg_SF}. The azimuthal symmetry curve fits tightly with the scatter plot as it is derived from a purely physical consideration.  

Nevertheless, the delimiting curve in $P_{GD}/\alpha_{GD}$ plane is distinct from that in the $H/\alpha$ plane. Firstly, the direction of the curve is reversed. This is because the physical depolarizers satisfying the Fry-Kattawar equation~\eqref{FWE} which also includes the coherent scatterers, all of which have a value of $P_{GD} = 1$. Secondly, the $P_{GD}$ has a physical lower bound for the physical depolarizers which is computed to be $0.25$. Thus, the zone with $P_{GD} < 0.25$ is never realized. This is achieved for the end point of the curve I evaluated for $m = 1$ whose Kennaugh matrix is given as,
\begin{equation*}
\mathbf{K}  =  \left[\begin{array}{cccc}
\frac{3}{2} & 0 & 0 & 0\\
0 & \frac{1}{2} & 0 & 0\\
0 & 0 & \frac{1}{2} & 0\\
0 & 0 & 0 & \frac{1}{2}
\end{array}\right],
\end{equation*}
for which the corresponding coherency matrix is given by
\begin{equation*}
\mathbf{T}  =  \left[\begin{array}{ccc}
1 & 0 & 0\\
0 & 1 & 0\\
0 & 0 & 1
\end{array}\right]. 
\end{equation*}
Thus, it is the case of degenerate eigenvalues with eigenvalue $1$ of multiplicity $3$. This also corresponds to the point of maximum entropy i.e., $H = 1$ in the $H/\alpha$ plane. This unique point in $P_{GD}/\alpha_{GD}$ scatter plot is characterized by $P_{GD} = 0.25$ and $\alpha_{GD} = \SI{90}{\degree} \times \cos^{-1}(1/\sqrt{3})\approx\SI{54.7356}{\degree}$. 

    \subsection{Interpretation of the Classes}
    
In our interpretation of the classes, the scattering behavior is determined by $\alpha_{GD}$, while $P_{GD}$ determines the purity/depolarizing nature of scattering. 
Table~\ref{tab:Class_ref} identifies the classes with the segments from $\alpha_{GD}$ and $P_{GD}$. 
Under this classification scheme, each pair starting from $(1,2), (3,4), (5,6)$ and $(7,8)$ belongs to the same zone in $\alpha_{GD}$ which is the proxy for the type of scattering as given in Table~\ref{tab:alpha_tau_angles}. 
The even-numbered class within each pair is purer or less depolarizing than the odd-numbered class. 

Classes $1$ and $2$ discriminate the sea from land. 
The vegetation mostly belongs to class $3$ because it is characterized as a distributed scatterer, and hence majorly depolarizing. 
The urban areas oriented about the radar LoS and those perpendicular are identified in class $5$ and $6$. 
Class $7$ is virtually absent because the corresponding $\alpha_{GD}$ segment is very narrow, i.e., $[\SI{80}{\degree}, \SI{90}{\degree}]$ and contains mostly pure scatterers viz., narrow dihedral and dihedral, hence, $P_{GD} > 0.5$.

\section{A Generic Scattering Power Factorization Framework}\label{Sec:SPFF}

In this section, we discuss a novel framework to obtain the component scattering powers using the order of dominance of similarity to known scattering models in PolSAR literature. 

\begin{figure}[!htbp]
	\centering
	\includegraphics[width=0.85\columnwidth]{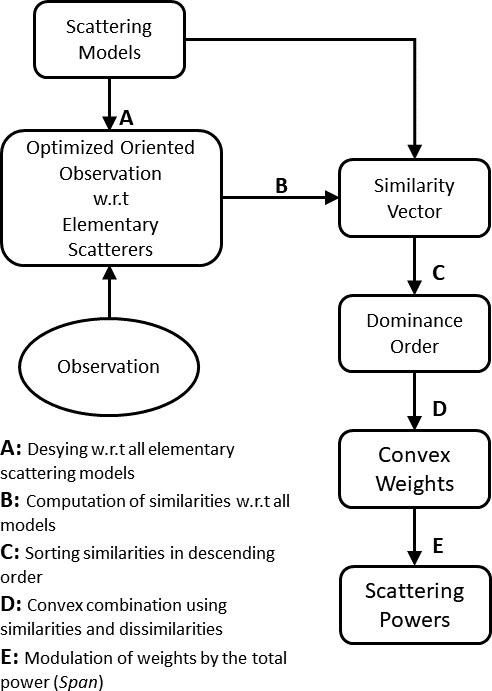}
	\caption{PolSAR Scattering Power Factorization Framework}
	\label{fig:Flow_gen}
\end{figure}

Fig.~\ref{fig:Flow_gen} outlines a generic scheme for scattering power factorization. 
It involves five key steps \textbf{A}\textendash\textbf{E} (as shown in the figure) that are applied to a PolSAR image in a pixel-by-pixel manner to obtain the scattering powers. 

This generic framework may be used for any similarity measure defined for the representation of polarimetric SAR data in the form of scattering matrix ($\mathbf{S}$), covariance or coherency matrix ($\mathbf{C}$ or $\mathbf{T}$) or Kennaugh matrix ($\mathbf{K}$). 
The data can be either coherent or incoherent. 
Moreover, it also provides the flexibility of using an arbitrary number of input scattering models to compare the observed data given in a particular representation.

In this study, we utilized this scattering power factorization framework, using the geodesic distance given in~\cite{Ratha_CD_2017,Ratha_SM_2017}. 
Within the flowchart, all intermediate products: the optimized radar line of sight (LoS) oriented observation w.r.t.\ the elementary scattering models, the similarity vector, the order of dominance of scattering components, the convex weights and the scattering powers can be used to infer various properties of a target.

Among the steps from \textbf{A}\textendash\textbf{E}, the transformation from similarity to convex weights takes place in step~\textbf{D}. 
In general, similarity is a quantity between $0$ and $1$, and its additive complement w.r.t. $1$ is the dissimilarity. Given, $n-$quantities $(x_1, x_2, \dots, x_n)$ between $0$ and $1$, a unique convex splitting of unity can be carried out in the following manner: 
\begin{eqnarray}
1 = x_1 + x_2(1-x_1) + \dots + x_n\cdots(1-x_2)(1-x_1) & \nonumber\\ 
+ (1-x_n)\cdots(1-x_2)(1-x_1) &
\end{eqnarray}
Except for the last term, all other terms are factored as one of the $n-$quantities while the rest of the factors are complements. The last term is the product of all the $n-$complements.  

\begin{figure}[!h]
	\centering
	\includegraphics[width=0.4\columnwidth]{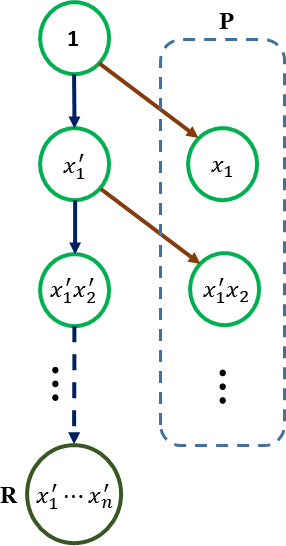}
	\caption{Convex splitting of unity}
	\label{fig:CS_Unity}
\end{figure}

Fig.~\ref{fig:CS_Unity} shows a diagram of the convex splitting of unity, denoting $x_i^\prime = (1-x_i)$. 
A single leaf node is produced by splitting a binary tree. 
Hence at the end of the split, one obtains the $n$ leaves contained in the dashed rectangle $\textbf{P}$, and an additional leaf $\textbf{R}$. 
It can be seen that the weights of the nodes in $\textbf{P}$ and $\textbf{R}$ add $1$ (root node).

\section{The $GD$ framework}\label{Sec:GDFramework}

\begin{figure*}[htb]
    \centering
    \includegraphics[width=0.75\textwidth]{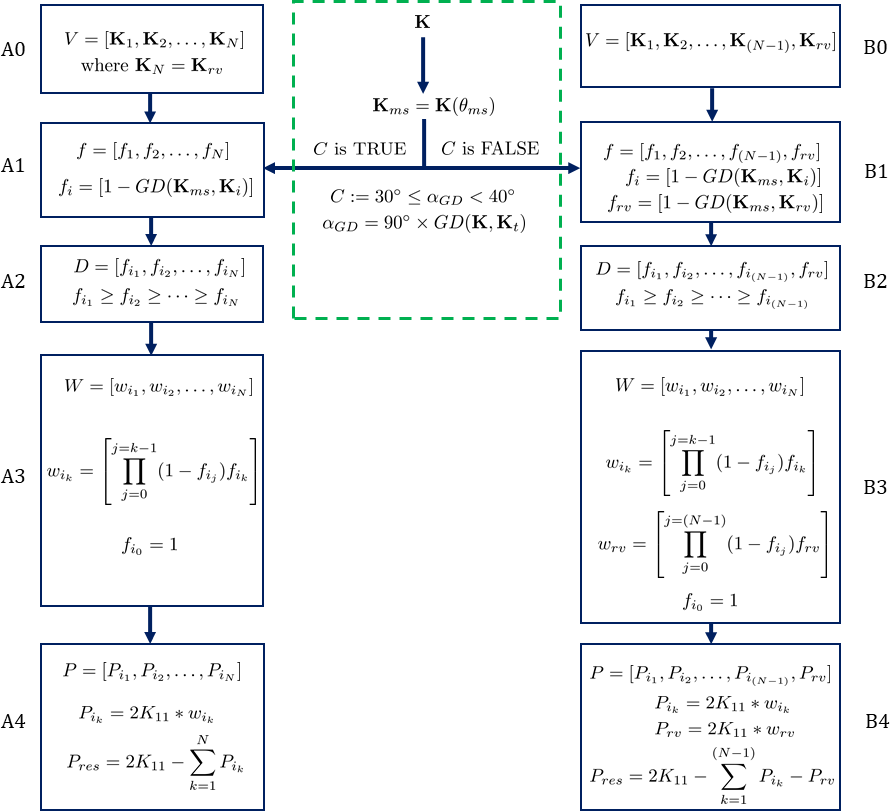}
    \caption{PolSAR Scattering Power Factorization Framework with volume model}
    \label{fig:Flow_CF_GD}
\end{figure*}

In this section, we utilize the framework outlined in Fig.~\ref{fig:Flow_gen} for multi-looked PolSAR images using a generalized volume scattering model~\cite{Antropov2011}. As a matter of fact, the preliminary yet impressive results in this direction were first obtained in~\cite{Ratha2019Fact}.
Fig.~\ref{fig:Flow_CF_GD} shows the scattering power factorization framework using $GD$ which is discussed here. 
In the flowchart, $\mathbf{K}_{1}$ to $\mathbf{K}_{(N-1)}$ are rank-$1$ elementary scatterers while $\mathbf{K}_{rv}$ given in~\eqref{eq:k_grv} denotes a volume scattering model. 
The parameter $\gamma = \langle |S_{HH}|^2\rangle/\langle|S_{VV}|^2\rangle$ i.e., the ratio of the pixel's co-polarized return intensities.

\begin{figure*}[hbt]
\begin{align}
\mathbf{K}_{rv} = \frac{1}{\frac{3(1+\gamma)}{4} - \frac{\sqrt{\gamma}}{6}}\begin{bmatrix}
\frac{3}{2}(1+\gamma) - \frac{\sqrt{\gamma}}{3} & \gamma - 1 & 0 & 0\\
\gamma - 1 & \frac{1}{2}(1+\gamma) + \frac{\sqrt{\gamma}}{3} & 0 & 0\\
0 & 0 & \frac{1}{2}(1+\gamma) + \frac{\sqrt{\gamma}}{3} & 0\\
0 & 0 & 0 & \frac{1}{2}(1+\gamma) - \sqrt{\gamma}
\end{bmatrix}
\label{eq:k_grv}
\end{align}
\end{figure*}

Initially, the observed Kennaugh matrix $\mathbf{K}$ is desyed to determine the maximal similarity with an elementary scatterer.
\begin{equation*}
GD(\mathbf{K}(\theta_{ms}),\mathbf{K}_j) = \min_{\theta,i}GD(\mathbf{K}(\theta),\mathbf{K}_i),
\end{equation*} 
where $i=1,2,\dots,(N-1)$ and $\theta \in [-\pi/8 \,, \pi/8]$. 
$\mathbf{K}(\theta)$ is the observed Kennaugh matrix in the $\theta$-rotated HV basis about the radar line of sight (LoS), earlier defined in~\eqref{eq:roll_K} of Sec.~\ref{Sec:RIP}.

Natural areas containing distributed scatterers show an $\alpha_{GD}$ value between \SIrange{30}{40}{\degree}. 
Thus, the condition $C := \SI{30}{\degree} \leq \alpha_{GD} < \SI{40}{\degree}$ is used for sorting pixels containing distributed targets from natural areas. 
Once the nature of the pixel is determined, a branch is assigned according to the truth value of $C$. 

Each branch consists of five labeled blocks: 
(0)~input of models, 
(1)~measurement of scattering similarities, 
(2)~determination of dominant similarity order, 
(3)~computation of convex weights with $GD$ and $(1-GD)$ as factors, and, finally, 
(4)~estimation of scattering powers by modulating the weights with $\text{Span} = 2K_{11}$ ($(1,1)^{th}$ element of $\mathbf{K}$). 

It is to be noted that the main difference between branches A and B is the position of $\mathbf{K}_{rv}$ in the similarity dominance vector $D$. 
In branch A, $[1-GD(\mathbf{K}_{ms},\mathbf{K}_{rv})]$ is allowed to position itself in $D$ according to its natural order in the array of similarities computed for each target. 
Whereas, in B it is forced to be the least dominant mechanism in the computing of the scattering power components. 
Finally, we compute the power components w.r.t. each of the input models, along with a residue power component $P_{res}$.
        
The rank-1 elementary scattering models for input are those of the trihedral ($t$), cylinder ($c$), narrow dihedral ($nd$), dihedral ($d$), left helix ($lh$) and right helix ($rh$). 
The restrictive volume model of Antropov et al.~\cite{Antropov2011} is used as the $\mathbf{K}_{rv}$ component. 
The scattering powers obtained in the output are grouped as shown in Table~\ref{tab:groupRGB} to produce a power composite RGB image.

\begin{table}[hbt]
    \centering
    \caption{Pseudocolor convention}
    \begin{tabular}{cccc}
        \toprule
        $P_{\text{odd}}$ & $P_{\text{rand}}$ & $P_{\text{even}}$ & $P_{\text{hlx}}$\\
        \cmidrule(lr){1-1}
        \cmidrule(lr){2-2}
        \cmidrule(lr){3-3}
        \cmidrule(lr){4-4}
        $P_{t} + P_{c}$ & $P_{rv} + P_{res}$ & $P_{nd} + P_{d}$ & $P_{lh} + P_{rh}$\\
        \cmidrule(lr){1-1}
        \cmidrule(lr){2-2}
        \cmidrule(lr){3-3}
        \cmidrule(lr){4-4}
        Blue & Green & Red & ---\\
        \bottomrule
    \end{tabular} 
    \label{tab:groupRGB}  
\end{table}

\section{Framework Results}\label{Sec:Results}

\begin{figure}[htb]
    \subfloat[Pauli RGB]{\includegraphics[width=0.39\columnwidth]{PauliRGB_fig_RS2}\label{subfig:PauliRGB_SF_RS2}}\qquad
    \subfloat[Y4R RGB\label{subfig:Y4R_RGB_SF_RS2}]{\includegraphics[width=0.39\columnwidth]{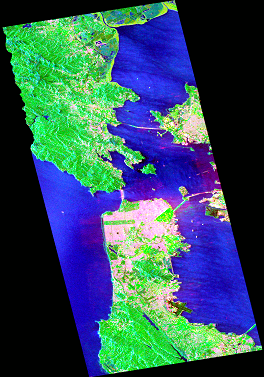}}\\
    \subfloat[RGB Composite]{\includegraphics[width=0.39\columnwidth]{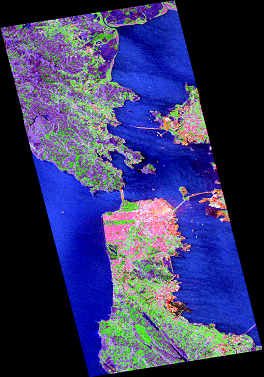}\label{subfig:CF_RGB_SF_RS2}}\hspace{0.25 em}
    \subfloat[Dominant Scattering Map\label{subfig:DS_SF_RS2}]{\includegraphics[width=0.58\columnwidth]{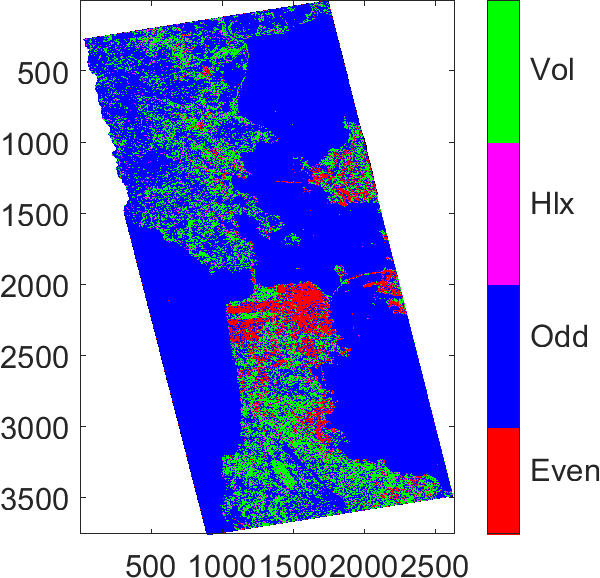}}\\
\caption{Pauli RGB and some output maps of framework for RS-2 C-Band San Francisco}
    \label{fig:Dominance_sf_rs2}
\end{figure}

\begin{figure}[htb]
    \subfloat[Pauli RGB]{\includegraphics[width=0.39\columnwidth]{PauliRGB_fig_ALOS2}\label{subfig:PauliRGB_SF_ALOS2}}\qquad
    \subfloat[Y4R RGB ]{\includegraphics[width=0.39\columnwidth]{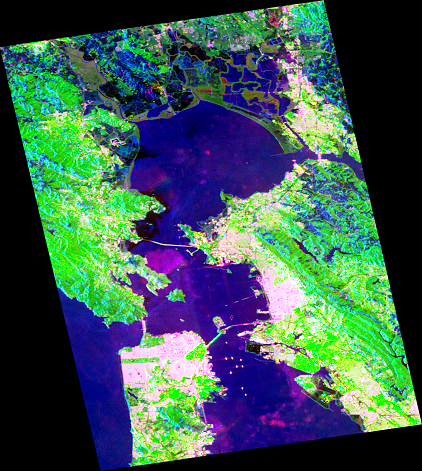}\label{subfig:Y4R_RGB_SF_ALOS2}}\\
    \subfloat[RGB Composite]{\includegraphics[width=0.39\columnwidth]{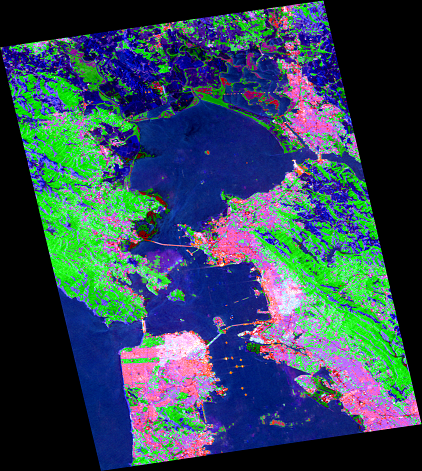}\label{subfig:CF_RGB_SF_ALOS2}}\hspace{0.25em}
    \subfloat[Dominant Scattering Map\label{subfig:DS_SF_ALOS2}]{\includegraphics[width=0.54\columnwidth]{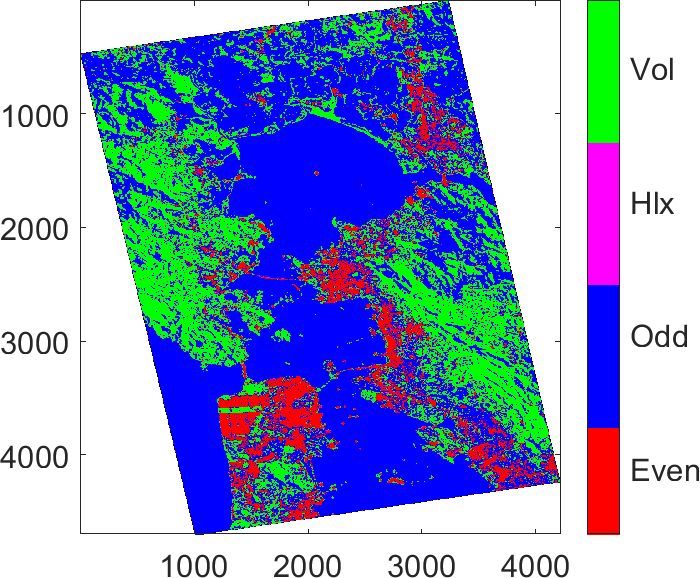}}\\
    \caption{Pauli RGB and some output maps of framework for ALOS-2 L-band San Francisco image}
    \label{fig:Dominance_sf_alos2}
\end{figure}

Figures~\ref{fig:Dominance_sf_rs2} and~\ref{fig:Dominance_sf_alos2} show the Pauli RGB, the Y4R~\cite{Yamaguchi2011} RGB, the SPFF RGB, and the dominant scattering class label from the framework for each pixel respectively for the San Francisco imaged by RS-2 and ALOS-2 sensors.


We can identify the South Market Area (SoMA)~\cite{Thirion2017XPol} in the SPFF RGB as an urban area which is present in both data sets. 
The Golden Gate park area is found to be a dominant volume scattering region while the sea surface is correctly identified as a dominant odd-bounce scattering zone.

A significant difference between the two scenes of San Francisco acquired by RS-2 and ALOS-2 sensors is the change in the dominant scattering label from odd-bounce to volume scattering for the region north of San Francisco across the iconic Golden Gate bridge. 
This is unlike the Pauli RGB and the Y4R RGB which show it as a dominant volume scattering zone. 
This area is typically a rugged terrain with sparse vegetation to dense forests. 
The images are acquired over a temporal gap of about 11 years. 
Additionally, the sensors are observing the area in different bands; hence the interaction with the targets is of different nature. 
An independent study using the $H/\alpha$ plot suggests that this region belongs to Zone~6 (medium entropy surface scatters) for the RS-2 data set, whereas it is Zone~5 (medium entropy vegetation scattering) for the ALOS-2 data set. 
As the framework uses the $\alpha_{GD}$ as a decisive parameter, this phenomenon is also clearly captured.  

\section{Conclusion}\label{Sec:Conclusion}

We have proposed a scattering power factorization framework and a few associated roll-invariant parameters for the analysis of PolSAR imagery. 
We have adopted a measure of similarity derived from the geodesic distance ($GD$) on the unit sphere in the space of $4 \times 4$ real matrices extended to the Kennaugh matrices. 
This $GD$ is shown to be bounded, measures only the scattering behavior, and is invariant to $\text{Span}$ scaling and to the orthogonal transformation of the $HV$ polarization basis. 
In the process, we have shown that the expression for the distance has convenient equivalent forms for the covariance/coherency and scattering matrices for multi-look and single-look data sets respectively. 

The framework is shown to be flexible concerning the addition of input scattering models. 
The convex splitting of unity helps in conserving the $\text{Span}$, thus providing a direct splitting of total power into components. 
It is shown to provide auxiliary information in the form of dominant scattering maps and roll-invariant parameters which are useful for classification studies. 
We also proposed three directly computable roll-invariant parameters, namely scattering type angle $\alpha_{GD}$, helicity $\tau_{GD}$, and depolarization index $P_{GD}$, within the scope of this framework; they were compared with their counterparts in the PolSAR literature. 
The proposed parameters are useful for effectively differentiating scattering behavior in a PolSAR scene.

Insights are obtained by analyzing sample scattering zones.
We showed, through a quantitative study, that the proposed parameters are expressive for classification.
These parameters are particularly useful for urban area mapping and land-cover classification, as vegetation and oriented urban areas are found to be separable by $\alpha_{GD}$ alone. 
The $\tau_{GD}$ is an apt parameter for separating land-ocean/ship-ocean scenes. 
Using these parameters in conjunction, we obtained a fast $P_{GD}/\alpha_{GD}$ classification scheme for PolSAR imagery which is capable in identifying different scattering zones within the scene.

In summary, we provided a dynamic and flexible framework for the analysis of PolSAR imagery, which is computationally easy to implement and fast to execute. 
This framework may be easily extended to accommodate an arbitrary number of scattering models. 
The proposed parameters may be used along with the complementary $\text{Span}$ information to obtain more scattering classes. 
In the future, it may also be used for the study of data sets in other polarimetric and bistatic configurations. 

\section*{Acknowledgment}
The first author would like to thank the Council of Scientific and Industrial Research (CSIR) for supporting his doctoral studies. 
He would also like to thank IFCPAR/CEFIPRA and Campus France for providing the opportunity to visit I.E.T.R.\ of Universiti\'e de Rennes1, France under the prestigious Raman-Charpak fellowship programme, during which a major part of this work was carried out.
A.\ C.\ Frery acknowledges support from CNPq and Fapeal.


\end{document}